\documentclass[onecolumn]{pasj01}
\usepackage{comment}

\begin{document} 
\Received{2017/07/10}
\Accepted{2017/10/25}

\title{FOREST Unbiased Galactic plane Imaging survey with the Nobeyama 45-m telescope (FUGIN) : Molecular clouds toward W33 ; possible evidence for a cloud-cloud collision triggering O star formation }

\author{Mikito \textsc{Kohno}\altaffilmark{1}$^{*}$%
}
\altaffiltext{1}{Department of Physics, Nagoya University, Furo-cho, Chikusa-ku, Nagoya, Aichi 464-8601, Japan}
\email{mikito@a.phys.nagoya-u.ac.jp}
\author{Kazufumi \textsc{Torii}\altaffilmark{2}}
\altaffiltext{2}{Nobeyama Radio Observatory, National Astronomical Observatory of Japan (NAOJ), National Institutes of Natural Sciences (NINS), 462-2, Nobeyama, Minamimaki, Minamisaku, Nagano 384-1305, Japan}
\author{Kengo \textsc{Tachihara}\altaffilmark{1}}
\author{Tomofumi \textsc{Umemoto}\altaffilmark{2,3}}
\author{Tetsuhiro \textsc{Minamidani}\altaffilmark{2,3}}
\altaffiltext{3}{Department of Astronomical Science, School of Physical Science, SOKENDAI (The Graduate University for Advanced Studies), 2-21-1, Osawa, Mitaka, Tokyo 181-8588, Japan}
\author{Atsushi \textsc{Nishimura}\altaffilmark{1}}
\author{Shinji \textsc{Fujita}\altaffilmark{1}}
\author{Mitsuhiro \textsc{Matsuo}\altaffilmark{2}}
\author{Mitsuyoshi \textsc{Yamagishi}\altaffilmark{4}}
\altaffiltext{4}{Institute of Space and Astronautical Science, Japan Aerospace Exploration Agency, Chuo-ku, Sagamihara  252-5210, Japan}
\author{Yuya \textsc{Tsuda}\altaffilmark{5}}
\altaffiltext{5}{Graduate School of Science and Engineering, Meisei University, 2-1-1 Hodokubo, Hino, Tokyo 191-0042, Japan}
\author{Mika \textsc{Kuriki}\altaffilmark{6}}
\altaffiltext{6}{Department of Physics, Graduate School of Pure and Applied Sciences, University of Tsukuba, 1-1-1 Ten-nodai, tsukuba, Ibaraki 305-8577, Japan}
\altaffiltext{7}{Institute for Advanced Research, Nagoya University, Furo-cho, Chikusa-ku, Nagoya 464-8601, Japan}
\author{Nario \textsc{Kuno}\altaffilmark{6}}
\author{Akio \textsc{Ohama}\altaffilmark{1}}
\author{Yusuke \textsc{Hattori}\altaffilmark{1}}
\author{Hidetoshi \textsc{Sano}\altaffilmark{1,7}}
\author{Hiroaki \textsc{Yamamoto}\altaffilmark{1}}
\author{Yasuo \textsc{Fukui}\altaffilmark{1,7}}



\maketitle

\begin{abstract}
{ We observed molecular clouds in the W33 high-mass star-forming region associated with
compact and extended H\,\emissiontype{II} regions using the NANTEN2 telescope as well as the Nobeyama 45-m
telescope in the $J=$1--0 transitions of $^{12}$CO, $^{13}$CO, and C$^{18}$O as a part
of the FOREST Unbiased Galactic plane Imaging survey with the
Nobeyama 45-m telescope (FUGIN) legacy survey. We detected three velocity
components at 35 km s$^{-1}$, 45 km s$^{-1}$, and 58 km s$^{-1}$. The 35 km s$^{-1}$ and 58 km s$^{-1}$
clouds are likely to be physically associated with W33 because of the
enhanced $^{12}$CO $J=$ 3--2 to $J=$1--0 intensity ratio as $R_{\rm 3-2/1-0} > 1.0$
due to the ultraviolet irradiation by OB stars, and morphological
correspondence between the distributions of molecular gas and the
infrared and radio continuum emissions excited by high-mass stars. 
The two clouds show complementary distributions around W33. The velocity separation 
is too large to be gravitationally bound, and yet not explained by
expanding motion by stellar feedback. Therefore, we discuss that a cloud-cloud
collision scenario likely explains the high-mass star formation in W33.}
\end{abstract}

\section{Introduction}
\subsection{High-mass star formation}
High-mass stars have a huge influence on the interstellar medium (ISM) and galactic evolution via stellar feedback and supernova explosions. { Feedback from high-mass stars is supposed to induce forming next generation stars as a sequential star formation (Elmegree {\&} Lada 1977; Lada 1987).
Supernova explosion  scatter heavy elements in the ISM that drives chemical evolution of galaxies.
However, formation mechanism of high-mass star is not clearly understood because it is difficult to achieve high mass accretion rate ($\sim10^{-3}$$M_{\odot}$ yr$^{-1}$, Wolfire {\&} Cassinelli 1987).
It is, therefore, {an} important issue to investigate what is the necessary condition for high-mass star formation in astrophysics.} 

{Based on} theoretical studies, two scenarios of high-mass star formation {are proposed}; the Core Accretion (monolithic collapse) and the Competitive Accretion (see Zinnecker {\&} Yorke 2007; Tan et al. 2014 {for reviews}). In the Core Accretion (monolithic collapse) model, high-mass stars are formed by the collapse of isolated gravitationally bound massive cores. It is a {similar process} to low-mass star formation {with} more massive aggregation (e.g., Nakano et al. 2000; Yorke {\&} Sonnhalter 2002; McKee {\&} Tan 2003; {Krumholz et al. 2007, 2009}; Hosokawa {\&} Omukai 2009a). On the other hand, {for} the Competitive Accretion model, high-mass stars are formed by growth of {low mass} protostellar seeds by mass accretion from surrounding gas (e.g., {Bonnell et al. 1997, 2001b, 2004}). One of the differences between {the} two models is the initial condition of {the natal cloud}. It is {therefore} important to observe high-mass star forming regions at very early stage of evolution which hold the initial condition of {high-mass star formation} in order to {verify} the theories, whereas such a {comparison} between theories and observations has not been {well made so far} and we do not have compelling evidence for each of the theory (Tan et al. 2014). Difficulties lie in the rareness of high mass star forming regions in the solar neighbor hood, and the short timescale of the feedback processes which is heavily mixed up with star formation processes.

\subsection{Cloud-Cloud Collisions as a trigger of high-mass star formation}
{During the past ten years, it is increasingly probable that a cloud-cloud collision plays an important role in high-mass star formation.}
In observational studies, a cloud-cloud collision was first reported in the star forming region NGC 1333 by Loren (1976). The Sagittarius B2 star forming region in the Galactic center was also suggested to {have} starburst triggered by a cloud-cloud collision based on {the} complementary distributions between {the cloud of} two velocity components (Hasegawa et al. 1994; Sato et al. 2000). 
In 2009, molecular observations with the NANTEN2 telescope showed two molecular clouds with different radial velocities toward a super star cluster Westerlund 2, and their distributions are interpreted as that the cluster formation was triggered by the collision of the two clouds (Furukawa et al. 2009). Further evidence for the physical association of the two molecular clouds with the cluster reinforced the interpretation (Ohama et al. 2010). Subsequently, three other super star clusters including 10-20 O stars are found to be associated with two molecular clouds with different velocities, and formation of O stars triggered by a cloud-cloud collision is likely to be a {common} process for clusters having more than 10 O stars (NGC 3603 {by} Fukui et al. 2014; RCW38 {by} Fukui et al. 2016; DBS[2003]179 {by} Kuwahara et al. in preparation). Among the 8 superstar clusters listed in the review article of {Portegies Zwart et al. (2010}), only three are known to be associated with localized nebulosities, indicating that the three are young and still associated with the {remnant of natal} molecular gas without heavy ionization. 
Additional possible cases of multiple O star formation triggered by a cloud-cloud collision are reported for M42 (Fukui et al. 2017b), NGC 6334-NGC 6357 (Fukui et al. 2017c), M17 (Nishimura et al. 2017a), W49A ({Miyawaki et al. 1986, 2009, Buckley \& Ward-Thompson 1996}), W51 (Okumura et al 2001; Kang et al. 2010 ; Fujita et al. in preperation) and R136 (Fukui et al. 2017a). 
In addition to above, many star forming regions and dense clumps in the Milky Way have been suggested to be triggered by a cloud-cloud collision, {LkH$\alpha$198 (Loren 1977); IRAS 19550+3248 (Koo et al. 1994); IRAS 2306+1451 (Vallee 1995)}; M20 (Torii et al. 2011, 2017a); RCW120 (Torii et al. 2015); N37 (Baug et al. 2016); GM 24 (Fukui et al. 2017d); M16 (Nishimura et al. 2017b); RCW34 (Hayashi et al. 2017); RCW36 (Sano et al. 2017a); RCW166 (Ohama et al. 2017a); S116, S117, S118 (Fukui et al. 2017e); { RCW79 (Ohama et al. 2017c)}; Sh2-48 (Torii et al. 2017); NGC 2024 (Ohama et al. 2017b); NGC 2068, NGC 2071 (Tsutsumi et al. 2017); { NGC 2359 (Sano et al. 2017b);} S87 (Xue \& Wu 2008); S87E, S88B, AFGL 5142, AFGL 5180 (Higuchi et al. 2010); G0.253+0.016 (Higuchi et al. 2014); {Circinus-E cloud (Shimoikura \& Dobashi 2011); Sh2-252 (Shimoikura et al. 2013)}; L1641-N (Nakamura et al. 2012); Serpens Main Cluster (Duarte-Cabral et al. 2011); Serpens South (Nakamura et al. 2014);  L1004E in the Cygnus OB 7 (Dobashi et al. 2014); G35.20-0.74 (Dewangan 2017a); {S235 (Dewangan, \& Ojha 2017b)}; L1188 (Gong et al. 2017); the Galactic Center 50 km s$^{-1}$ molecular cloud (Tsuboi et al. 2015); N159 West (Fukui et al. 2015) and N159 East (Saigo et al. 2017) in the Large Magellanic Cloud.

In theoretical studies, hydrodynamical numerical {simulation} of a cloud-cloud collision was first carried out by Stone (1970a, 1970b). Habe {and} Ohta (1992) made numerical simulations of head-on collisions for two clouds of different sizes. They showed that gravitationally unstable cores are created by compression between the two clouds ({ see also, Anathpindika 2010, Takahira et al. 2014, 2017, {Shima et al. 2017}}). Balfour et al. (2015, 2017) also presented numerical simulation of head-on and non-head-on a cloud-cloud collision. {In addition, the magneto-hydrodynamical simulations (MHD) simulation of giant molecular cloud (GMC) collision was carried out by several authors (e.g., Wu et al. 2015, 2017a, 2017b; Christie et al. 2017; Bisbas et al. 2017; {Li et al. 2017}). They showed GMC collision enhanced star formation rate and efficiency (Wu et al. 2017b).} Inoue {and} Fukui (2013) studied the interface layer of the colliding clouds by three-dimensional MHD simulations, and showed formation of massive molecular cores, which likely lead to form high-mass protostars gaining high mass accretion rate helped by amplified turbulence and magnetic fields via supersonic collision ({ see also Inoue et al. 2017). Kobayashi et al. (2017a, 2017b) discussed the evolution of GMC mass functions including cloud-cloud collisions.} 
In the global scale numerical simulations, cloud-cloud collisions is important mechanism of star formation in the Galaxy (e.g., Tan 2000, Tasker \& Tan 2009, {Fujimoto et al. 2014a, 2014b}, Dobbs et al. 2015, {Li 2017}). 

These observational and theoretical { studies} suggest that a cloud-cloud collision is a promising mechanism of massive star formation, whereas there still remain large number of massive star forming regions where cloud-cloud collisions has not been investigated well.

\subsection{High-mass star forming region W33}
W33 is a high-mass star forming region, which was first cataloged by the 1390 MHz radio continuum survey (Westerhout 1958), extending for 10 pc$\times$10 pc centered on $(l,b) \sim$(\timeform{12.8D}, \timeform{-0.2D}). The parallactic distance of W33 was measured as 2.4 kpc based on the water maser observations by Immer et al. (2013) indicating that W33 is located in the Scutum spiral arm in the Milky Way. Figure 1 shows a three color composite image of the {\it Spitzer} space telescope observations (GLIMPSE : Benjamin et al. 2003, MIPSGAL : Carey et al. 2009), where blue, green, and red correspond to the 3.6 $\mu$m, 8 $\mu$m, and 24 $\mu$m emissions, respectively. The contours in Figure 1 indicate the MAGPIS 90 cm radio continuum emission (Helfand et al. 2006).

W33 harbors many star forming clumps, OB stars, and H\,\emissiontype{II} regions. { There are six dust clumps (W33 Main, W33 A, W33 B, W33 Main1, W33 A1, and W33 B1) as shown in the { pink} contours in Figure 1 obtained by the Atacama Pathfinder Experiment (APEX) Telescope Large Area Survey of the GALaxy (ATLASGAL) 870 $\mu$m survey (Schuller et al. 2009; Contreras et al. 2013; Urquhart et al. 2014).  The sizes and masses of the dust clumps are listed together with their physical properties in Table 1. {These six dust clumps are suggested to be on different evolutional stages indicated by their spectral properties and associations with radio continuum 
sources (Immer et al. 2014)}. W33 Main includes a (compact) H II region detected by the radio continuum observations (Hachick \& Ho 1983, Ho et al. 1986), while W33 A and W33 B harbor hot cores by the chemical line survey. W33 Main1, W33 A1, and W33 B1 are identified as high-mass proto-stellar objects (Immer et al. 2014).}
Radio observations by Haschick \& Ho (1983) revealed the presence of an obscured (proto-)cluster which includes a number of high-mass stars with spectral types from O7.5 to B1.5 by assuming a distance of 4 kpc. Associations of the maser sources also support nature of the young massive stellar objects, i.e., water and methanol masers in W33 Main, W33 A, and W33 B, and OH masers in W33 A and W33 B (Immer et al. 2013; Menten et al. 1986; Caswell 1998; Colom et al. 2015). The Submillimeter Array (SMA) high-resolution observations showed that W33 Main harbors multiple ultra-compact H\,\emissiontype{II} regions and three high density clumps (W33 Main-A, W33 Main-B1, W33 Main-B2) embedded in a dense gas envelope detected in C$_2$H (Jiang et al. 2015).  W33 A has been studied based on the observations in various wavelengths (Gibb et al. 2000; Roueff et al. 2006; Davies et al. 2010; de Wit et al. 2007, 2010). Galv\'an-Madrid et al. (2010) detected high-velocity gas associated with an outflow in the CO $J =$ 2--1 transition by SMA 230 GHz band observations in W33 A. They suggested that star formation activity in W33 A was triggered by filamentary convergent gas flows from two different velocity components. Recently, Maud et al. (2017) carried out Atacama Large Milimeter/Submilimeter Array (ALMA) observations in Band 6 and Band 7 with 500 AU scale resolution toward W33 A. The authors found spiral and filamentary structures around the central massive young stellar object in W33 A.

H\,\emissiontype{II} regions in W33 is shown in the radio continuum emissions shown in Figure 1 as white contours, where the names of the H\,\emissiontype{II} regions cataloged by the radio recombination line survey (G012.745-00.153, Downes et al. 1980; Lockman 1989) and Wide-Field Infrared Survey Explorer (WISE) satellite (Anderson et al. 2014, 2015); G012.692-00.251, G012.820-00.238, G012.884-00.237, and G012.907-00.277, are labeled.

Messineo et al. (2015) identified many high-mass stars in W33 based on the near infrared K-band observations using Spectrograph for INtegral Field Observations in the Near Infrared (SINFONI) on the Very Large Telescope (VLT). { Positions} of the OB stars identified by Messineo et al. (2015) are indicated as circles on the three-color image of the {\it Spitzer} observations in Figure 1, where the circles colored in orange indicate the O-type stars, while those in white indicate early B-type stars. Two O4-6 are distributed in G012.745-00.153, and  Messineo et al. (2015) discussed that the two O stars are evolved to be (supper-)giant. The authors discussed the ages of these two O stars as less than 6 Myr based on a stellar model by Ekst\"{o}m et al. (2012) or 2-4 Myr by comparing the O4-6 supergiants in the Arches cluster in the Galactic center (Martins et al. 2008), which show similar $K$-band spectra with the present two O stars in W33.

CO rotational transition line observations at the millimeter wavelength covering the entire W33 region were carried out using the Five College Radio Observatory (FCRAO) 14 m telescope and Millimeter Wave Observatory (MWO) 5 m telescope by Goldsmith and Mao (1983), revealing that W33 has complicated velocity structures, whereas detailed spatial and velocity distributions of molecular gas remained unclear. Observations of the H$_2$CO absorption, and radio recombination lines indicate that W33 Main has a velocity at $\sim$ 35 km s$^{-1}$, while W33 B is at $\sim$60 km s$^{-1}$ (Gardner \& Whiteoak 1972; Bieging et al. 1978). Immer et al. (2013) detected {{ water}} maser emission at a velocity of $\sim$35 km s$^{-1}$ toward W33 Main and W33 A and $\sim$60 km s$^{-1}$ toward W33 B, finding that these two velocity components are at the same annual parallaxial distance of 2.4 kpc. The authors concluded that W33 is a single star forming region in spite of the large velocity difference between the two velocity components, $\sim$ 25 km s$^{-1}$, although the origin of the two velocity components is still ambiguous.

In order to reveal the detailed spatial and velocity distributions of the molecular gas in W33, we carried out high-resolution observations in the $J=$ 1--0 transition of $^{12}$CO, $^{13}$CO, and C$^{18}$O using the NANTEN2 and Nobeyama 45-m telescopes. This paper organizes as follows; section 2 describes observations. Section 3 gives observational results and comparison with archive datasets. Section 4 discusses a cloud-cloud collision scenario of W33. Section 5 concluded this paper.

\section{Dataset}
\subsection{NANTEN2 $^{12}$CO $J =$ 1--0 Observations}
The NANTEN2 4m millimeter/sub-millimeter telescope { of Nagoya University} situated in Chile was used {to} observe a large area of W33 in $^{12}$CO $J=$ 1--0 emission {with} the On-The-Fly (OTF) mode from 2012 May to 2012 December.  The half-power beam width (HPBW) {is \timeform{2.7'}} at 115 GHz. This corresponds to 1.9 pc at the distance of 2.4 kpc. A 4 K cooled superconductor-insulator-superconductor (SIS) mixer receiver provided a typical system temperature of $\sim$ 250 K in Double Side Band (DSB), and {a 16384 channel digital spectrometer with} a bandwidth and  resolution of 1 GHz and 61 kHz, which corresponds to 2600 km s$^{-1}$ {and} 0.16 km s$^{-1}$,  respectively, at 115GHz. We smoothed the obtained data {to} a velocity resolution of 1 km s$^{-1}$ and angular resolution of \timeform{200"}. The pointing accuracy was confirmed to be better than \timeform{15"} with the daily observations of the Sun and IRC+10216. 
{ We used chopper wheel method to calibrate the antenna temperature ($T_a^*$) (Penzias \& Burrus 1973, Ulich \& Haas 1976, Kuter \& Ulich 1981).
The absolute intensity fluctuation was calibrated by daily observations of IRAS16293-2422 [$\alpha_{\rm J2000} = \timeform{16h32m23.3s} , \delta_{\rm J2000} = \timeform{-24D28'39. 2''}$] and the intensity scale was converted into $T_{\rm mb}$ scale by assuming its peak $T_{\rm mb}= 18$ K (Ridge et al. 2006). The intensity uncertainty of NANTEN2 datasets is $< 20 \%$. The typical root-mean-square (rms) noise level is $\sim$ 0.6 K with a \timeform{200"} smoothing data and velocity resolution of 1.0 km s$^{-1}$.} 

\subsection{Nobeyama 45-m Telescope $^{12}$CO $J =$ 1--0,  $^{13}$CO $J =$ 1--0, C$^{18}$O $J =$ 1--0 Observations}
Detailed CO $J=$1--0 data around W33 were obtained by using the Nobeyama 45-m telescope in Nobeyama Radio Observatory (NRO).
The half-power beam width (HPBW) {is} \timeform{14"} at 115 GHz{, and \timeform{15"} at 110 GHz}. This corresponds to 0.2 pc at the distance of 2.4 kpc. We simultaneously observed in $^{12}$CO, $^{13}$CO, {and} C$^{18}$O as a part of FUGIN (FOREST Unbiased Galactic plane Imaging survey with the Nobeyama 45-m telescope; Umemoto et al. 2017; { Minamidani et al. 2015}) legacy survey with the OTF mode (Sawada et al. 2008) { from 2014 March to 2015 May}. 
FOREST (FOur-beam REceiver System on the 45-m Telescope) is {a four-beam}, dual-polarization, and two sideband (2SB) receiver, {providing} a typical system temperature of $\sim$ 150 K in $^{13}$CO $J=$ 1--0, {and} $\sim$ 250 K in $^{12}$CO $J=$ 1--0 ({ Minamidani et al. 2016}).
We used SAM45 (Spectral Analysis Machine for the 45-m telescope: Kuno et al. 2011), that is a FX type digital spectrometer same as the ALMA ACA Correlator (Kamazaki et al. 2012). It {has} 4096 channels {with} a bandwidth and resolution of 1 GHz and 244.14 kHz, which corresponds to 2600 km s$^{-1}$ {and} 1.3 km s$^{-1}$, respectively, at 115 GHz. We smoothed the obtained data {to the} angular resolution of \timeform{30"}. The typical pointing accuracy was confirmed to be better than \timeform{3"} observing SiO maser sources, { such as V VX Sgr [$\alpha_{\rm B1950} = \timeform{18h05m02.959s} , \delta_{\rm B1950} = \timeform{-22D13'55.58"}$], for the observing run toward W33 every hour with the 40 GHz HEMT receiver named H40}. 
{The intensity variation was calibrated with daily observations of the W51D {[$\alpha_{\rm B1950} = \timeform{19h21m22.2s} , \delta_{\rm B1950} = \timeform{14D25'17.0"}$]}. { We used chopper wheel method to convert antenna temperature ($T_a^*$) (Penzias, \& Burrus 1973, Ulich \& Haas 1976, Kuter \& Ulich 1981).}
 { The data in the antenna temperature ($T_a^*$) scale {was} converted into main beam temperature ($T_{\rm mb}$) as $T_{\rm mb}  = T_a^* / \eta_{\rm mb}$, where the main beam efficiency ($\eta_{\rm mb}$) of 0.43 for $^{12}$CO and 0.45 for  $^{13}$CO and C$^{18}$O (Umemoto et al. 2017, Minamidani et al. 2016).} 
{ The typical rms noise levels after intensity calibration ( $T_{\rm mb}$ scale) are $\sim 0.5$ K, $\sim 0.2$ K, and $\sim 0.2$ K, for $^{12}$CO, $^{13}$CO, and C$^{18}$O $J=$ 1--0, respectively}.
{ The intensity variation of $^{12}$CO, $^{13}$CO, and C$^{18}$O $J=$ 1--0 are 10--20 \%, 10\%, and 10\%, respectively. }
The FUGIN project overview paper gives more detailed information (Umemoto et al. 2017). { We summarized the observational parameters of  NANTEN2 and Nobeyama 45-m { datasets} as Table 2.}

\subsection{{Archive dataset}}
{We use the $^{12}$CO $J=$3--2 archival data of {CO High Resolution Survey (COHRS)} obtained 
with JCMT (James Clark Maxwell Telescope: {Dempsey et al. 2013}). 
The spatial and velocity resolutions are 1.0 km s$^{-1}$ and \timeform{16"}{,} respectively. We smoothed the data to {an} angular resolution of \timeform{30"}. The data in the antenna temperature ($T_a^*$) scale {was} converted into main beam temperature ($T_{\rm mb}$) with the equation of $T_{\rm mb}  = T_a^* / \eta_{\rm mb}$, where the main beam efficiency ($\eta_{\rm mb}$) of 0.61 {was} adopted by planet observations with the uncertainty of about 10\% - 15\% ({Dempsey et al 2013;} Buckle et al. 2009).
{ The typical rms noise in the $T_{\rm mb}$ scale was $\sim 0.2$ K for $^{12}$CO $J=$ 3--2. }

We use the following datasets to compare with the CO data. 
i.e., near and mid-infrared data from {\it Spitzer} space telescope (GLIMPSE in 3.6 $\mu$m and 8.0 $\mu$m, Benjamin et al. 2003, MIPSGAL in 24 $\mu$m, Carey et al. 2009),
 the 20 cm and 90 cm free-free radio continuum data from MAGPIS (A Multi-Array Galactic Plane Imaging Survey ; Helfand et al. 2006) observed with the Very Large Array (VLA); and the H\,\emissiontype{I} 21cm emission data from SGPS II (Southern Galactic Plane Survey : McClure-Griffiths et al. 2005) observed with the Australia Telescope Compact Array (ATCA) and the Parkes Radio Telescope. The angular resolution of the H\,\emissiontype{I} data is {$\sim \timeform{3.3'}$} and their spectrum resolutions is 0.8 km s$^{-1}$.}

\section{Results}
\subsection{Distribution and properties of molecular gas}

Figure 2(a) shows the longitude-velocity diagram of the NANTEN2 $^{12}$CO $J =$ 1--0 data for a large area including W33. There are four distinct velocity components toward W33 as depicted by arrows.  { Goldsmith and Mao (1983) mentioned with CO observations that association of the 5--25 km s$^{-1}$ velocity component indicated as black arrow with W33 is not clear}, as it is likely to be located in the Sagittarius arms (Reid et al. 2016). In this study, we therefore focus on the velocity components at 35 km s$^{-1}$, 45 km s$^{-1}$, and 58 km s$^{-1}$ as candidate clouds associated with W33. We hereafter refer these three clouds as “the 35 km s$^{-1}$ cloud'', “the 45 km s$^{-1}$ cloud'', and “the 58 km s$^{-1}$ cloud''. These three clouds appear to be connected with each others in the velocity space (Figure 2(a)). The 35 km s$^{-1}$ cloud is continuously distributed along the Galactic longitude for the present region, while the CO emissions in the other two clouds at higher velocities are enhanced especially toward W33 { (dashed box in Figure 2(a))}. Figure 2(b) shows the longitude-velocity diagram of the three velocity clouds of W33 using the FUGIN $^{12}$CO $J=$ 1--0 data, which shows the detailed velocity distribution of the gas at high spatial resolution. The three velocity clouds can be separately identified at this spatial scale in Figure 2(b).

Figures 3 shows the spatial distributions of the three clouds in the four CO { lines}. The left, center, and right columns show the 30 km s$^{-1}$, 45 km s$^{-1}$, and 58 km s$^{-1}$ clouds, respectively, where the contours indicate the MAGPIS 90 cm radio continuum emission, and the positions of { the dust clumps} and OB stars identified by Messineo et al. (2015) are depicted by crosses and circles, respectively. We also present the velocity channel maps of the four CO transitions obtained with Nobeyama 45-m and JCMT in Figures 14-17 in the appendix as supplements. The CO emission in the 35 km s$^{-1}$ cloud is enhanced at the corresponding region of W33 in all the transitions. W33 Main shows the brightest CO emission in this region.

We found counterparts of W33 A, { W33 Main}, and W33 B1 in the 35 km s$^{-1}$ cloud,  in the C$^{18}$O emission ({ Figure 3 (g)}). { Figure 4 shows C$^{18}$O distributions of each dust clumps. The C$^{18}$O molecular clump properties {discussed in Section 3.2.}} Among these sources, W33 Main and W33 A are associated with CO molecular outflows as discussed later in Section 3.6. The 35 km s$^{-1}$ cloud also shows morphological anti-correlations with the radio continuum emissions in H\,\emissiontype{II} regions G012.745-00.153 and G012.820-00.238. The CO emission in G012.745-00.153 is enhanced at the eastern $(l,b)=(12.78,-0.18)$ and southern rim $(l,b)=(12.74,-0.18)$ of the H\,\emissiontype{II} region, showing steep intensity gradient in the $^{12}$CO emissions, while G012.820-00.238 is surrounded by molecular gas, especially in the $^{13}$CO and C$^{18}$O emissions. W33 Main is sandwiched by these two H\,\emissiontype{II} regions.
The 45 km s$^{-1}$ cloud has diffuse CO emission extended over the present region. The compact emissions at W33 Main and W33 A correspond to the wing features of the outflows (see { Section 3.6} for details), and are thus not related to the 45 km s$^{-1}$ cloud. 
Molecular gas in the 58 km s$^{-1}$ cloud is separated into the northern and southern components relative to W33 and the central part corresponding to W33 is weak in the CO emission. There are several clumpy structures embedded at the northern rim of the southern component, which are clearly seen in the $^{12}$CO emissions, and these clumps show spatial correlations with radio continuum emissions from the H\,\emissiontype{II} regions G012.745-00.153 and G012.692-00.251 as well as W33 B. In the C$^{18}$O map in Figure 3(i) { and Figure 4(d)}, W33 B is associated with the strong CO peak. There are several other clumpy molecular structures at the interspace between the northern and southern components of the 58 km s$^{-1}$ cloud, forming an arc-like molecular structure which looks surrounding W33. { The size of arc-like structure is roughly estimated to be $\sim$ 7 pc. {On the other hand, clear associations of molecular clumps with W33 A1 and W33 Main1 are not recognized.}}

We derived the column densities and masses of the three velocity clouds using the $^{12}$CO integrated intensity maps shown in Figures 3(a)-(c), where we defied the individual clouds by drawing contours at {{ $5 \sigma$ noise levels {in the integrated intensity of 8 K km s$^{-1}$ for the velocity interval of 10 km s$^{-1}$}}}. By assuming a $X$(CO) factor of $2\times10^{20}$ (K km s$^{-1})^{-1}$ cm$^{-2}$ (Strong et al. 1988), we estimated the mean column densities of the 35 km s$^{-1}$, 45 km s$^{-1}$, and 58 km s$^{-1}$ clouds as {{ $1.7 \times 10^{22}$ cm$^{-2}$, $1.7 \times 10^{22}$ cm$^{-2}$ and $6.2 \times 10^{21}$ cm$^{-2}$, respectively, with the total molecular masses derived as $1.1 \times 10^5 M_{\odot}, 1.0 \times 10^5 M_{\odot}$, and $3.8 \times 10^4 M_{\odot}$.  The uncertainty of mass estimation using X-factor is about $\pm$ 30 \% (Bolatto et al. 2013). Lin et al. (2016) derived the mean column densities as $2.5\times10^{22}$ cm$^{-2}$ using the infrared dust emission data obtained by {\it Herschel}, which is consistent with our estimate.}}

\subsection{C$^{18}$O molecular clump properties}
{  We define C$^{18}$O molecular clumps using the following procedures in order to investigate the physical properties of dense molecular gas {belonging to the 35 km s$^{-1}$ and 58 km s$^{-1}$ clouds} corresponding to the dust clumps.
\begin{enumerate}
\item Search for a peak integrated intensity toward {the} six dust clumps.
\item Define a clump boundary as the half level of its peak integrated intensity. 
\item If the area enclosed by the boundary have multiple peaks, define the boundary as a contour of the 75 \% level of its peak integrated intensity.
\end{enumerate}
We identified four molecular clumps corresponding to the dust clumps of W33 Main, W33 A, and W33 B1 associated with the 35 km s$^{-1}$ cloud (Figure 4(a), 4(b), and 4(c)), and W33 B associated with the 58 km s$^{-1}$ cloud (Figure 4(d)), while we could not define the boundary of the molecular clump toward W33 Main1 and W33 A1, {because W33 A1 does not have { an intensity} peak toward the dust clump and W33 Main1 can not be separated from the extended feature (Figure 4(e) and (f))}.

Then, we derived the physical parameters assuming the Local Thermal Equilibrium (LTE) using the following procedures (Wilson et al. 2009). 

\begin{enumerate}
\item  Derive {the} excitation temperature ($T_{\rm ex}$) assuming that the $^{12}$ CO $J=$ 1--0 transition {line} is optically thick and $T_{\rm ex}$ is uniform throughout the molecular clump by following the equation from the $^{12}$ CO peak intensity ($T_{\rm mb}(\rm ^{12}CO peak)$) at the peak position of the clump
\begin{eqnarray}
T_{\rm ex} &=& 5.5 \bigg/ \ln \left(1+ {5.5 \over T_{\rm mb}(\rm ^{12}CO peak) + 0.82 }\right).
\label{eq:75}
\end{eqnarray}
\item Estimate the optical depth of the C$^{18}$O emission ($\tau_{18}$) at each pixel and velocity channel from {the} C$^{18}$O brightness temperature ($T_{\rm mb}(v)$) at velocity $v$ {as}
\begin{eqnarray}
\tau_{18} (v) &=& -\ln \left[1-{T_{\rm mb}(v) \over 5.3} \left\{ {1 \over \exp({5.3 \over T_{\rm ex}})-1}-0.17 \right\}^{-1} \right].
\label{eq:75}
\end{eqnarray}
\item Calculate the C$^{18}$O column density ($N ({\rm C^{18}O})$) at each pixel summing up the {quantities} of all $v$ channels {whose resolution is} 1 km s$^{-1}$.
\begin{eqnarray}
N ({\rm C^{18}O}) &=& 2.4 \times 10^{14} \times \sum_v {T_{{\rm ex}} \tau_{18} (v) \Delta v \over 1-\exp \left(-{5.3 \over T_{\rm ex}} \right) } 
\label{eq:75}
\end{eqnarray}

\item $N ({\rm C^{18}O})$) is converted into H$_2$ column density ($N(\rm H_2)$) assuming the following conversion formula derived from the Ophiuchus molecular cloud (Frerking et al. 1982). 
\begin{eqnarray}
N ({\rm H_2}) = \left[ {N (\rm C^{18}O) \over 1.7 \times 10^{14}} + 3.9 \right] \times 10^{21}
\label{eq:75}
\end{eqnarray}
\item Estimate the size {($r$) of each clump assuming that clumps are spherical shapes using the equation of 
\begin{eqnarray}
r = \sqrt{{S \over \pi}}=\sqrt{D^2\Omega \over \pi},
\label{eq:75}
\end{eqnarray}
where $S$ is the area of the clump enclosed by the boundary contour, $D$ is the distance to W33, and $\Omega$ is the solid angle of the clump.
\item {Calculate the mass of each clump as}
\begin{eqnarray}
M =  \mu m_{\rm H} D^2 \Omega \sum N(\rm H_2),
\label{eq:75}
\end{eqnarray}
}
where $\mu$ is the mean molecular weight 2.8, $m_{\rm H}$ is the proton mass, and the summation is performed over the each clump within {the boundary.}
\item The averaged number density of hydrogen molecules is calculated as 
\begin{eqnarray}
n({\rm H_2}) ={ 3 M \over 4 \pi r^3 \mu m_{\rm H}  }
\label{eq:75}
\end{eqnarray}
by assuming that the clumps have a uniform density.
\item The virial mass is estimated by the equation of 
\begin{eqnarray}
M_{\rm vir} ={ 5 r \Delta v^2 \over 8 (\ln2) G  }=210 \left({r \over [\rm pc]}\right) \left({\Delta V_{\rm comp} \over [\rm km\ s^{-1}]}\right)^2
\label{eq:75}
\end{eqnarray}
from the virial theorem using the clump size ($r$) and the Full Width of Half Maximum (FWHM) of the composite spectrum $\Delta V_{\rm comp}$ in the clump {estimated by} a single Gaussian fitting. 

\end{enumerate}

Table 3 presents the physical properties of {the} molecular clumps. The typical values of {the peak column density, mean column density, mass, size, hydrogen molecules number density, and virial mass are $N({\rm H_2})_{\rm peak} \sim 10^{23}\ {\rm cm^{-2}}, N({\rm H_2})_{\rm mean} \sim 10^{22}\ {\rm cm^{-2}}$, $M_{\rm clump} \sim 10^3$ -- $10^4\ M_{\odot}$, $r \sim$ \ {0.4 -- 1 pc, $n(\rm H_2) \sim 10^4$ -- $10^5\ $cm$^{-3}$, and $M_{\rm vir} \sim 10^2 $-- $10^3 \ M_{\odot}$}. The peak and {mean} column densities is roughly consistent with those calculated from the dust continuum observations (Immer et al. 2014).}

\subsection{CO $J=$ 3--2/$J=$ 1--0 intensity ratio}

Figures 5(a), (b), and (c) show distributions of the $^{12}$CO $J=$ 3--2/$J=$ 1--0 intensity ratio ($R_{\rm 3-2/1-0}$) using \timeform{50"} smoothing data (rms $\sim$ 0.26 K) in the 35 km s$^{-1}$, 45 km s$^{-1}$, and 58 km s$^{-1}$ clouds, respectively. { We adopted the clipping level as $ 8 \sigma$.} Line intensity ratios between different $J$ levels of CO provide useful diagnostics to investigate physical association of molecular gas with high-mass stars, {{ as these depend on the gas kinematic temperature ($T_{\rm k}$) and number density of hydrogen ($n({\rm H_2})$) following the Large Velocity Gradient (LVG) model. (e.g., Goldreich \& Kwan 1974).}} Figure 5(a) shows that the 35 km s$^{-1}$ cloud has a high $R_{\rm 3-2/1-0}$ of $>$1.0 around the central part of W33 which includes W33 Main, W33 Main1, and W33 B1. The 45 km s$^{-1}$ cloud shows lower $R_{\rm 3-2/1-0}$ of 0.4--0.5 except for the corresponding parts of W33 Main and W33 A, where $R_{\rm 3-2/1-0}$ is locally elevated up to 0.8--1.2. As discussed in the next subsection, the high $R_{\rm 3-2/1-0}$ in W33 Main and W33 A in the 45 km s$^{-1}$ cloud are due to the outflows emitted from the protostars embeded in the 35 km s$^{-1}$ cloud. The 58 km s$^{-1}$ cloud has high $R_{\rm 3-2/1-0}$ of 1.2 in the arc-like structure which seems to surround W33, while the gas outside the arc-like structure shows low $R_{\rm 3-2/1-0}$ of $\sim$0.4. We also present the velocity channel map of the $R_{\rm 3-2/1-0}$ in Figure 18 in the appendix. The cause of the high $R_{\rm 3-2/1-0}$ of gas in the 35 km s$^{-1}$ and 58 km s$^{-1}$ clouds can be understood as heating by the high-mass stars in W33. The high $R_{\rm 3-2/1-0}$ of molecular gas in the 35 km s$^{-1}$ and 58 km s$^{-1}$ clouds { can be understood as high kinematic temperature heated by the high-mass stars in W33. Therefore, the 35 km s$^{-1}$ and 58 km s$^{-1}$ clouds indicate physical associations of W33, whereas association of the 45 km s$^{-1}$ is elusive.}

\subsection{Comparisons of the 35 km s$^{-1}$ and 58 km s$^{-1}$ clouds with infrared data}
We here focus on the 35 km s$^{-1}$ and 58 km s$^{-1}$ clouds, as these are most likely associated with W33. Figure 6 demonstrates comparisons between the $^{12}$CO $J =$ 3--2 and infrared emissions, where the {\it Spitzer} 8 $\mu$m and 24 $\mu$m emissions are shown in grayscale in Figures 6(a) and (b), respectively. { The 8 $\mu$m emission trace thermal emission from hot dust plus emission from Polycyclic Aromatic Hydrocarbon (PAH), which is distributed in the Photo Dissociation Region (PDR) {(e.g., Churchwell et al. 2004)}.  The 24 $\mu$m emission also traces warm dust grain heated by bright high-mass stars in the H\,\emissiontype{II} region {(e.g, Carey et al. 2009, Deharveng et al. 2010).}
}The red and blue contours show the 35 km s$^{-1}$ and 58 km s$^{-1}$ clouds, respectively. The arc-like structure in the 58 km s$^{-1}$ cloud surrounds { the strong emission part of the 35 km s$^{-1}$ cloud}, showing complementary distributions between these two. The 8 $\mu$m emission, which is bright at the south-eastern rim of the H\,\emissiontype{II} region G012.745-00.153 $(l,b)\sim(12.75, -0.18)$, coincides with the steep intensity gradient of the CO emission in the 35 km s$^{-1}$ cloud. This lends more credence to association between the 35 km s$^{-1}$ cloud and G012.745-00.153.

The 24 $\mu$m emission shown in Figure 6(b) is attributed to the warm dust grains and thus can be used to probe the region where the heating by the high-mass stars in W33 is efficient. The 24 $\mu$m emission in W33 is enhanced at { the dust clumps} and the H\,\emissiontype{II} regions. Especially, G012.745-00.153 and G012.820-00.238 as well as W33 Main show bright  24 $\mu$m emissions. The arc-like structure in the 58 km s$^{-1}$ cloud overall traces the outline of the 24 $\mu$m distribution. This suggests that the high $R_{\rm 3-2/1-0}$ of gas in the arc-like structure shown in Figure 5(c) is due to interaction with the star forming regions which are spatially correlated with the arc-like structure.  

The morphological correlations of the 58 km s$^{-1}$ cloud with infrared emissions, as well as the high $R_{\rm 3-2/1-0}$ of gas in the 58 km s$^{-1}$ cloud, strongly suggests that not only the 35 km s$^{-1}$ cloud but also the 58 km s$^{-1}$ cloud are physically associated with W33 despite of a large velocity separation between the two clouds, $\sim$23 km s$^{-1}$. This is consistent with the previous studies by Gardner \& Whiteoak (1972), Bieging et al. (1978), and Immer et al. (2013) based on the observations of H$_2$CO absorption, radio recombination line, and water masers, in which Immer et al. (2013) discussed that the two velocity components are at the same annual parallax distance of 2.4 kpc.

On the other hand, association of the 45 km s$^{-1}$ cloud at the intermediate velocity range between the 35 km s$^{-1}$ and 58 km s$^{-1}$ clouds with W33 is not evident in the present dataset.

\subsection{{Comparison with the H\,\emissiontype{I} 21 cm line}}
We analyzed the H\,\emissiontype{I} 21 cm line data in W33 using the archival data obtained in the Southern Galactic Plane Survey (SGPS) project (McClure-Griffiths et al. 2005). Figure 7 shows the integrated intensity distributions of the three velocity clouds of the H\,\emissiontype{I} data. The contours show the MAGPIS 20 cm radio continuum emission, where the original image was spatially smoothed to have the same resolution as that of the SGPS H\,\emissiontype{I} data. In Figure 7(a) the H\,\emissiontype{I} emission in the 35 km s$^{-1}$ cloud shows strong intensity depression at W33 Main where the 20 cm continuum emission is strong. The depression can be confirmed in the H\,\emissiontype{I} 21 cm spectra shown in { Figure 8(a)}. The H\,\emissiontype{I} profile has negative intensities at the velocity range of the 35 km s$^{-1}$ cloud as indicated by the shade. This indicates absorption against the background continuum source, implying that W33 Main is distributed at the inside or at the rear side of the 35 km s$^{-1}$ cloud. The H\,\emissiontype{I} maps of the 45 km s$^{-1}$ and 58 km s$^{-1}$ clouds in Figures 7(b) and (c) show weak intensity depressions toward G012.745-00.153 and W33 B, whereas it is not clear toward W33 Main. The H\,\emissiontype{I} spectra toward G012.745-00.153 shown in { Figure 8(b) and (c)} show absorption features at the velocity ranges where the $^{12}$CO emission appears, i.e., 40--50 km s$^{-1}$ in { Figure 8(b)} and at 50 -- 60 km s$^{-1}$ in { Figure 8(c)}. {{ We suggested that these absorptions { are} the unrelated foreground components overlapping on the line of sight. }}

\subsection{Molecular outflows}

We identified bipolar outflows in the CO emissions toward W33 Main and W33 A. Figures 9 and 10 show the CO spectra and spatial distributions of the outflow lobes in W33 Main and W33 A, respectively. The outflows in both of W33 Main and W33 A have total velocity widths as large as 40--50 km s$^{-1}$. The systemic velocities of these outflows are estimated to be $\sim$35 km s$^{-1}$ using the optically thin C$^{18}$O $J =$ 1--0 emission. W33 A has been reported to be associated with outflows (Davies et al. 2010, de Wit et al. 2010, Galv\'an-Madrid et al. 2010). As the blue and red lobes in Figures 9 and 10 are not fully resolved in the present CO data due to the limitation of the spatial resolution, and we here estimate the lengths of the outflow lobes to be 0.5 pc, { defined by contours at half of the peak intensity level.} { If we assume inclination angle of \timeform{45D}}, the dynamical timescale ($t_{\rm dyn}$) of W33 Main and W33 A are calculated from the maximum velocity ($V_{\rm max} \sim$ 23 km s$^{-1}$) and size ($r$) to be { $ 0.5/23 \times \sqrt{2} \sim 3 \times 10^4$ yr, which is consistent with previous JCMT $^{12}$CO and $^{13}$CO $J=$ 3--2 observation results (Maud et al. 2015).}

\section{Discussion}
The results of the present observations and analyses are summarized as follows:
\begin{enumerate}
\item  We identified three molecular clouds toward W33 at $\sim$35 km s$^{-1}$, $\sim$45 km s$^{-1}$, and $\sim$58 km s$^{-1}$, using the NANTEN2, FUGIN CO $J=$ 1--0 , { and JCMT $^{12}$CO $J=$ 3--2} data. The total molecular masses of the three clouds are derived as {{ $1.1 \times 10^5 M_{\odot}, 1.0 \times 10^5 M_{\odot}$, and $3.8 \times 10^4 M_{\odot}$, respectively.}} 
\item  The 35 km s$^{-1}$ cloud is spatially coincident with W33. Our CO data revealed a spatial correlation of the 35 km s$^{-1}$ cloud with the { dust clumps of W33 Main, W33 A, W33 B1, and the H\,\emissiontype{II} regions G012.745-00.153 and G012.820-00.238, having $R_{\rm 3-2/1-0}$ of higher than 1.0.} 
{{ { A strong} absorption feature in the H\,\emissiontype{I} 21 cm line is seen in the velocity range of the 35 km s$^{-1}$ cloud. Therefore, W33 Main is located inside or behind the 35 km s$^{-1}$ cloud. }}
\item The 45 km s$^{-1}$ cloud shows diffuse CO emission extended for the present region, and its association with W33 is not clear in terms of spatial correlation. 
\item In the 58 km s$^{-1}$ cloud the present CO dataset revealed an arc-like structure having a size of $\sim$7 pc. It shows complementary distributions with the 35 km s$^{-1}$ cloud and the infrared images along the line-of-sight, having $R_{\rm 3-2/1-0}$ of higher than 1.0. These observational properties suggest association of the arc-like structure with the dust clump W33 B and the H\,\emissiontype{II} regions G012.745-00.153 and G012.692-00.251 distributed at the southern part of W33. 
\item We identified two bipolar molecular outflows toward W33 Main and W33 A. The full velocity widths of the outflows are as large as 40--50 km s$^{-1}$. The dynamical timescales of the outflows can be estimated to be about $\sim 3 \times 10^4$ yr. 
\end{enumerate}

In this section, we discuss the origin of the observed properties of the multiple velocity clouds in W33 over 23 km s$^{-1}$ and these relationships with the high-mass star formation in W33.

\subsection{Gravitationally binding of the multiple velocity clouds}
Based on the proper motion measurements, Immer et al. (2014) argued if W33 A and W33 B are gravitationally bound to W33 Main, and the authors concluded that W33 A and W33 B are not gravitationally bound to W33 Main, as the total speed of W33 A and W33 B are larger than the derived escape velocity.

In addition to their calculations, we here test dynamical binding of the 35 km s$^{-1}$ and 58 km s$^{-1}$ clouds, as these contain the high-mass star forming regions other than W33 Main, W33 A, and W33 B. {{ If we assume that the two clouds are separated by 10 pc in space, the same order of the cloud size as a rough order-of-magnitude estimation,}} and by 23 km s$^{-1}$ in velocity, the total mass required to gravitationally bind these two clouds can be calculated as $M = {rv^2 \over 2G}= 6 \times 10^5\ M_{\odot}$. This figure is {{ by a factor 4 larger than the total molecular mass associated with W33, $1.5 \times 10^5 M_{\odot}$ }}, calculated in { Section 3.1}, indicating that the co-existence of the two velocity clouds in W33 can not be interpreted as the gravitationally bound system.

\subsection{Expanding motion driven by feedback from high-mass stars}
Another idea to interpret the multiple velocity clouds is expanding motion driven by feedback from high-mass stars. If we assume a spherical expansion of gas being interacted with feedback, it would display a ring-like velocity distribution both in a spatial map and a position-velocity diagram (e.g., see Figure 8 of Torii et al. 2015). { The expanding velocity ($v_{\rm exp}$) of an H\,\emissiontype{II} region is limited by the sound speed of the ionized gas confined in the H\,\emissiontype{II} region, which corresponds to $v_{\rm exp} \sim$12 km s$^{-1}$ with the electron temperature of 10000 K (e.g., Ward-Thompson, \& Whitworth 2011). For the present case in W33, the arc-like structure in the 58 km s$^{-1}$ cloud and velocity separation of 23 km s$^{-1}$ look consistent with this assumption. However, the longitude-velocity diagram presented in Figure 2 shows that both of the 35 km s$^{-1}$ and 58 km s$^{-1}$ clouds are separated having uniform velocities for a large area including W33. These velocity distributions in the position-velocity diagram are inconsistent with the prediction from the expansion assumption.} Dale et al. (2013) discussed that contribution of stellar wind to expansion of neutral medium is relatively minor compared with expansion of H\,\emissiontype{II} region at gas density at $\sim10^4$ cm$^{-1}$. We thus conclude that the co-location of the multiple velocity clouds in W33 over 23 km s$^{-1}$ was not formed by the feedback from the high-mass stars in W33.

\subsection{The cloud-cloud collision model}
We here postulate a cloud-cloud collision scenario as an alternative idea to interpret the multiple velocity clouds associated with W33. In a collision between two clouds with different sizes (Habe and Ohta 1992), a smaller cloud drives into a larger cloud, forming a cavity on the larger cloud, and a dense gas layer is formed at the interface of the collision, which corresponds to the bottom of the cavity, by strong compression of gas. High-mass stars are formed in this dense gas layer. In addition, a thin layer with turbulent gas is formed at the interface between the larger cloud and the dense layer, which has intermediate velocities between the larger cloud and the dense layer. This thin, turbulent layer is observed as “broad bridge features" in a position-velocity diagram. which is the emission between two velocity peaks with intermediate intensities (e.g., Torii et al. in prep.; Torii et al. 2017a; Haworth et al. 2015a; Haworth et al. 2015b). The turbulent gas is replenished as long as the collision continues. 

Another observational signature of cloud-cloud collisions is “complementary distribution between two velocity clouds" (Fukui et al. 2017b; Torii et al. 2017a). For an observer viewing angle parallel to the collisional axis so that the two clouds are spatially coincident, the observer can see an anti-correlated or complementary distribution between the two clouds separated in velocity, as the larger cloud with a cavity displays a ring-like gas distribution on the sky. Fukui et al. (2017b) pointed out that, if the viewing angle has an inclination relative to the colliding axis, complementary distribution has a spatial offset, which is proportional to the projected travel distance of the collision.

Gas distribution in the position-velocity diagram including broad bridge feature depends on various parameters such as cloud shapes, density contrast of the clouds, initial relative velocity between the clouds, etc. In Figures 11 and 12, we show schematics of two extreme cases of a cloud-cloud collision and cartoons of the corresponding position-velocity diagram, based on the discussions by Haworth et al. (2015a) and Haworth et al (2015b), in which the authors post-processed the model data of cloud-cloud collisions calculated by Takahira et al. (2014). 

\begin{enumerate}
\item  If the smaller cloud has a gas density much smaller than the larger cloud, which is the case shown in Figure 11, the moving velocity of the dense layer is immediately decreased owing to the momentum conservation, and the collision halts in the middle of the larger cloud. In this case, the entirety of the smaller cloud undergoes collision quite quickly, streaming into the dense layer. The cartoon position-velocity diagram in the middle of the collision in Figure 11 show two velocity peaks separated by broad bridge features having intermediate intensities, whereas there is only one velocity peak after the collision is stopped.
\item  On the other hand, if the smaller cloud has a density much higher than the larger cloud, deceleration of the dense layer is quite small, and the smaller cloud punches the larger cloud within a short time (Figure 12). In this case, although gas distribution in the position-velocity diagram during the collision is similar to that in the case 1 in the 0th order approximation, gas configuration after the collision finishes is quite different from the case 1. In the position-velocity diagram at the final stage, as only some small fraction of the smaller cloud undergoes the collision, the two colliding clouds are separated in velocity without broad bridge features, and the larger cloud shows an intensity depression at the positions corresponding to the smaller cloud. In this case, spatially complementary distribution is an important diagnostic to investigate cloud-cloud collisions.
\end{enumerate}

\subsection{Cloud-cloud collision in W33}
We here discuss an application of the cloud-cloud collision model to the observed clouds in W33. We propose a scenario that a collision between a smaller cloud having a size of $\sim$7 pc (which corresponds to the 35 km s$^{-1}$ cloud) and a larger cloud with a size of $\sim$17 pc (the 58 km s$^{-1}$ cloud) occurred in this region. In this scenario, it can be interpreted that the arc-like structure in the 58 km s$^{-1}$ cloud which shows complementary distribution with the 35 km s$^{-1}$ cloud was created through the collision. 
As shown in the position-velocity diagram in Figure 2(b), these two clouds are separately in velocity. In addition, since the 35 km s$^{-1}$ cloud has a similar velocity compared with its surroundings outside the sightline of W33 as shown in Figure 2(b), deceleration of the colliding velocity was as small as less than 4--5 km s$^{-1}$, which corresponds to the linewidth of the 35 km s$^{-1}$ cloud. 
These signatures suggest the case 2 collision discussed in the previous subsection in Figure 12. {{ The mean density of the {35 km s$^{-1}$ and 58 km s$^{-1}$ cloud} are roughly estimated to be $1.7 \times 10^{22}\ {\rm cm}^{-2}/ 7\ {\rm pc} \sim\ 2\times 10^3\ {\rm cm}^{-3}$ and $6.2 \times 10^{21}\ {\rm cm}^{-2}/ 17\ {\rm pc} \sim 400\ {\rm cm}^{-3}$ {from the $^{12}$CO $J=$1--0 dataset}, respectively.}} { In addition, the density of molecular clump W33 Main in the center of cavity associated with the 35 km s$^{-1}$ cloud and W33 B associated with the 58 km s$^{-1}$ cloud is $\sim 10^5$ cm$^{-3}$ and $\sim 10^4$ cm$^{-3}$ {from the C$^{18}$O $J=$1--0 dataset}, respectively. Therefore we assumed the density of the smaller cloud is much higher than that in the larger cloud. }
Based on the schematic of the final stage of the collision in Figure 12, the timescale of the collision can be estimated to be 17--(17+7) pc / 23 km s$^{-1} \sim$ 0.7--1.0 Myr, depending on the lengths of the dense layer and the smaller cloud in this stage.

\subsection{High-mass star formation in W33}
In the previous subsection, we discussed that the observed gas properties in W33 can be interpreted with the cloud-cloud collision model. We here discuss relationships between the collision and the high-mass star formation in W33. 

As already introduced, W33 contains high-mass star forming regions in various evolutionary stages. The six dust clumps in W33, i.e., W33 Main, W33 A, W33 B, W33 Main1, W33 A1, and W33 B1, have ages younger than the estimated timescale of the cloud-cloud collision, 0.7--1.0 Myr, { because of the dynamical time scale of the outflow in W33 Main and W33 A are $\sim 10^4$yr. We suggest} that these sources were formed in the dense layer created though the collision between the 35 km s$^{-1}$ and 58 km s$^{-1}$ clouds. It is possible that continuous compression and accumulation of gas by the collision led a mass accretion rate as high as $10^{-4}$--$10^{-3}$ $M_\odot$ yr$^{-1}$ at theses objects, which satisfies the condition of high-mass star formation (e.g, Wolfire {\&} Cassinelli 1987; Hosokawa et al. 2010; McKee {\&} Tan 2003).

In addition to these sources, there are several extended H\,\emissiontype{II} regions in W33.  Messineo et al. (2015) identified four new O stars in W33 as indicated in Figure 1 by the orange circles. Among them, two O4--6I stars are distributed in G012.745-00.153, while one O4--6I star is in G012.820-00.238. The authors discussed that these three O stars have ages less than 6 Myr based on a stellar model by Ekst\"{o}m et al. (2012) or 2--4 Myr by comparing the O4--6 supergiants in the Arches cluster in the Galactic center (Martins et al. 2008), which show similar $K$-band spectra with the present O stars in W33. The estimate of 2--4 Myr is larger than our estimate of 0.7--1.0 Myr on the timescale of the cloud-cloud collision in W33.

In order to investigate the formation timescales of these O stars in a different way, we here estimate the formation timescale of the H\,\emissiontype{II} regions G012.745-00.153 and G012.820-00.238, that are excited by thees O4--6I stars. In Figure 13 evolutionary tracks of the H\,\emissiontype{II} region radius $r_{\rm HII}$ are plotted for different initial densities, based on the analytical model of the D-type expansion by Spitzer (1978). Here the Lymann continuum photons $N_{\rm Ly}$ of $10^{49.55}$--$10^{49.93}$ s$^{-1}$, which correspond to a O6I--O4I star, are assumed (Panagia 1973). The horizontal dashed lines in Figures 13(a) and (b) indicate the radii of G012.745-00.153 and G012.820-00.238 measured by the 90 cm data, respectively.

In Figure 13(a), in which the case for G012.745-00.153 is presented, the evolution timescales which correspond to the measured $r_{\rm HII}$ are about 0.2 Myr, 0.7 Myr, and 2 Myr for initial densities of $10^3$ cm$^{-3}$, $10^4$ cm$^{-4}$, and $10^5$ cm$^{-3}$, respectively, suggesting that a quite high density of $\sim10^5$ cm$^{-3}$ is required to be consistent with the age of 2--4 Myr estimated by Messineo et al. (2015). However, in the case of G012.820-00.238 shown in Figure 13(b), even with a quite high density of $10^5$ cm$^{-3}$, the derived timescale is as short as 0.5 Myr, and it is inconsistent with the estimate by Messineo et al. (2015) based on the comparison with the Arches cluster in the Galactic center, although their another estimate of less than 6 Myr, which is based on the stellar model by Ekst\"{o}m et al. (2012), is consistent with the present results in Figure 13. 

If we tentatively assume that the ages of G012.745-00.153 and G012.820-00.238 are 0.7 Myr and 0.2 Myr, respectively, in which a reasonable initial gas density of $10^4$ cm$^{-3}$ is assumed, these figures are within the estimated timescale of the cloud-cloud collision in W33, 0.7--1 Myr, and it is therefore possible that formation of these O stars in G012.745-00.153 and G012.820-00.238 were occurred in the dense layer formed through the cloud-cloud collision between the 35 km s$^{-1}$ and 58 km s$^{-1}$ clouds.

\section{Conclusions}
The conclusions of the present study are summarized as follows.

\begin{enumerate}
\item We carried out large scale CO observations using the NANTEN2 and Nobeyama 45-m telescopes toward the galactic high-mass star forming region W33. Our dataset identified three velocity clouds at 35 km s$^{-1}$, 45 km s$^{-1}$, and 58 km s$^{-1}$ toward W33. { The 35 km s$^{-1}$ cloud spatially coincide with W33, showing spatial correlations with the dust clumps W33 Main, W33 A, and W33 B1} and the H\,\emissiontype{II} regions G012.745-00.153 and G012.820-00.238, while the 58 km s$^{-1}$ cloud, which has an arc-like structure surrounding the 35 km s$^{-1}$ cloud and thus W33, shows associations with the dust clump W33 B and the H\,\emissiontype{II} regions G012.745-00.153, and G012.692-00.251. The 45 km s$^{-1}$ shows diffuse and extended CO emission throughout the corresponding region of W33, although its association with W33 is not clear. The total molecular masses of the three clouds, i.e., the 35 km s$^{-1}$, 45 km s$^{-1}$, and 58 km s$^{-1}$ clouds, are {{ $1.1 \times 10^5 M_{\odot}, 1.0 \times 10^5 M_{\odot}$, and $3.0 \times 10^4 M_{\odot}$, respectively.}}

\item We analyzed the intensity ratio between $^{12}$CO $J=$ 3--2 and $^{12}$CO $J=$ 1--0 in these three velocity clouds, finding that the 35 km s$^{-1}$ and 58 km s$^{-1}$ clouds have $R_{\rm 3-2/1-0}$ of higher than 1.0 in the vicinity of W33, whereas the 45 km s$^{-1}$ cloud have low $R_{\rm 3-2/1-0}$ of $\sim$0.4 throughout the analyzed region. The cause of the high $R_{\rm 3-2/1-0}$ of gas in the 35 km s$^{-1}$ and 58 km s$^{-1}$ clouds can be understood as heating by the high-mass stars in W33, suggesting physical associations of these two velocity clouds with W33. 

\item {{ In the H\,\emissiontype{I} 21 cm line profiles, strong absorption features are detected at 35 km s$^{-1}$ toward W33 Main. This result indicates that W33 Main is located inside or behind the 35 km s$^{-1}$ cloud}}.

\item We identified two CO bipolar outflows toward W33 Main and W33 A. The full velocity widths of the outflows are as large as 40--50 km s$^{-1}$. The dynamical timescales of the outflows can be estimated to be about $\sim 3 \times 10^4$ yr. 

\item In order to interpret the co-location of the two velocity components at W33, we discussed three possibilities, i.e., gravitationally bound system, expanding motion by feedback from high-mass stars in W33, and cloud-cloud collision model. We found that neither of the former two assumptions can explain the observed signatures of the molecular gas.

\item Finally we proposed a cloud-cloud collision model to interpret the multiple velocity clouds in W33. We assumed a collision of  two molecular clouds with different sizes at a relative velocity of 23 km s$^{-1}$. In this model, the arc-like structure in the 58 km s$^{-1}$ cloud can be reasonably explained as a cavity created by the collision. The uniform velocity distributions of the 35 km s$^{-1}$ and 58 km s$^{-1}$ clouds in the position-velocity diagram suggest that the gas density of the smaller cloud, which corresponds to the 35 km s$^{-1}$ cloud, is much higher than that in the large cloud, the 58 km s$^{-1}$ cloud. This is consistent with the observations. The timescale of the collision was estimated to be 0.7--1.0 Myr. We also discussed that it is possible for the high-mass star forming regions in W33 to have been formed in the dense gas layer created through the collision.

\end{enumerate}

\section*{Acknowledgements}
{ We are grateful to Misaki Hanaoka for a useful discussion.}
NANTEN2 is an international collaboration of ten universities: Nagoya University, Osaka Prefecture University, University of Cologne, University of Bonn, Seoul National University, University of Chile, University of New South Wales, Macquarie University, University of Sydney, and Zurich Technical University.
The Nobeyama 45-m radio telescope is operated by Nobeyama Radio Observatory, a branch of National Astronomical Observatory of Japan. Data analysis was carried out on the open use data analysis computer system at the Astronomy Data Center, ADC, of the National Astronomical Observatory of Japan.

The work is financially supported by a Grant-in-Aid for Scientific Research (KAKENHI, No. 15K17607, 15H05694) from MEXT (the Ministry of Education, Culture, Sports, Science and Technology of Japan) and JSPS (Japan Soxiety for the Promotion of Science).

\clearpage

\begin{table*}
\tbl{Observational properties of dust clumps in W33}{
\begin{tabular}{cccccccccc}
\hline
\multicolumn{1}{c}{Source} & $\ell$ & $b$ & Size$^{(1)}$ & Mass$^{(1)}$ & Evolutional Stage$^{(2)}$ &  $V_{\rm LSR, H_2O}^{(3)}$ & $V_{\rm LSR, H_2CO}^{(4)}$  \\
&[degree]& [degree] & [pc] & [$M_{\odot}$] & &[km s$^{-1}$]& [km s$^{-1}$] \\
\hline
W33 Main\footnotemark[]  & 12.804 & $-0.200$ & 0.25 & $4 \times 10^3$& (Compact) H\,\emissiontype{II} region & 34  &$\sim 35$\\
W33 A\footnotemark[] &12.907 & $-0.259$ & 0.15 & $3 \times 10^3$ & Hot core   & 35 & $\sim 35$\\
W33 B\footnotemark[] & 12.679 & $-0.182$ & 0.1 & $2 \times 10^3$ & Hot core & 59 &$\sim 60$\\
W33 Main1\footnotemark[] & 12.852 & $-0.225$ & 0.1 & $5 \times 10^2$ & High-mass protostellar object &  -- & -- \\
W33 A1\footnotemark[] & 12.857  & $-0.273$ & 0.1 & $4 \times 10^2$ & High-mass protostellar object &   -- & --  \\
W33 B1\footnotemark[] &  12.719  & $-0.217$ & 0.1 & $2 \times 10^2$ & High-mass protostellar object&   -- & --  \\
\hline
\end{tabular}}\label{tab:first}
\begin{tabnote}
\footnotesize References: (1) Immer et al. (2014), (2) Haschick \& Ho (1983); Immer et al. (2014), (3) Immer et al. (2013), (4) Gardner \& Whiteoak (1972)   \\
\end{tabnote}
\end{table*}

\begin{table*}
\tbl{Observational properties of NANTEN2 and FUGIN datasets.}{
\begin{tabular}{cccccccccc}
\hline
\multicolumn{1}{c}{Telescope} & Line & HPBW  & Effective  &  Velocity & RMS noise$^2$  \\
& & & Beam Size & Resolution & level \\
\hline
NANTEN2 &$^{12}$CO $J=$ 1--0 \footnotemark[]  &  \timeform{160"} & \timeform{180"} & 0.16 km s$^{-1}$ & $\sim 0.6$ K  \\
Nobeyama 45-m$^1$  &$^{12}$CO $J=$ 1--0\footnotemark[] &  \timeform{14"} &  \timeform{20"} & 1.3 km s$^{-1}$& $\sim 0.5$ K   \\
&$^{13}$CO $J=$ 1--0\footnotemark[] & \timeform{15"} &  \timeform{21"} & 1.3  km s$^{-1}$& $\sim 0.2$ K \\
&C$^{18}$O $J=$ 1--0\footnotemark[] &  \timeform{15"}   &   \timeform{21"} & 1.3 km s$^{-1}$& $\sim 0.2$ K \\
\hline
\end{tabular}}\label{tab:first}
\begin{tabnote}
\footnotemark[1] Reference: Umemoto et al. (2017)\\
\footnotemark[2] The value of rms noise level is after smoothing datasets. \\
\end{tabnote}
\end{table*}

\begin{table*}[h]
\tbl{Physical parameters of the C$^{18}$O molecular clumps in W33. }{
\begin{tabular}{cccccccccccccc}
\hline
\multicolumn{1}{c}{Clump}  & $V_{\rm LSR}$ &$T_{\rm ex}$ & $\tau_{18}$ &  $N(\rm H_2)_{\rm peak}$ & Size & $\Delta V_{\rm comp}$ &  $N(\rm H_2)_{\rm mean}$  &$M_{\rm clump}$ &$n(\rm H_2)$& $M_{\rm vir}$\\
&[km s$^{-1}$] &  [K] & &[cm$^{-2}$] & [pc] & [km s$^{-1}$] & [cm$^{-2}$] & [$M_{\odot}$]  & [cm$^{-3}$] & [$M_{\odot}$]\\
(1) & (2) &  (3) & (4) & (5) & (6) & (7) & (8) & (9)  & (10) & (11)\\
\hline
W33 Main  & 35.5 & 34 & 0.11  & $6.0 \times 10^{23}$ & 0.60  & 5.5 & 5.1 $\times 10^{22}$   & $9.5 \times 10^3$ & $1.5 \times 10^5$& 3.8 $\times 10^3$\\
W33 A\footnotemark[$*$]  & 36.5 & 18 & 0.15  & $2.6 \times 10^{23}$& {1.0} & 5.5 & 3.7 $\times 10^{22} $ &  $1.4 \times 10^4$ &{$4.8 \times 10^4$}& {6.4 $\times 10^3 $}\\
W33 B  & 56.5 & 17 & 0.096  & $1.9 \times 10^{23}$ & 0.81 & 6.4 & 2.1 $\times 10^{22} $ &  $5.1 \times 10^3$&$3.3 \times 10^4$& 6.9 $\times 10^3$\\
W33 Main1  & 36.5 & 19 & 0.096 & $2.0 \times 10^{23}$& -- & -- & -- &  -- & --  & -- \\
W33 A1 & 36.5 & 18 & 0.12 &  $2.1 \times 10^{23}$ & --  & -- & --& -- &-- &--\\
W33 B1\footnotemark[$*$] & 35.5 & 23 & 0.13 & $1.2 \times 10^{23}$ & {0.41} & 3.1 &1.3 $\times 10^{22} $&  $ 1.9 \times 10^3$&{$9.2 \times 10^4$}& {8.3 $\times 10^2$}\\
\hline
\end{tabular}}\label{tab:first}
\begin{tabnote}
\footnotemark[$*$] {Defined by the 75\% level boundary.} \\
\footnotesize Note. Columns: (1) Name. (2) C$^{18}$O peak velocity (3) Excitation temperature of $^{12}$CO $J=$ 1--0 peak intensity. (4) Optical depth of peak position. (5) Peak column density of the C$^{18}$O clump. (6) Size of the C$^{18}$O clump. (7) Line width of the composite profile obtained by averaging spectra in the C$^{18}$O clump. (8) Average column density of the C$^{18}$O clump. (9) The Molecular mass within the clump size with the assumption of LTE. (10) Density of the clump with the assumption of sphere. (11) Virial mass of the clump.  W33 Main1 and W33 A1 show only peak parameters because we could not define the size of clump.\\
\end{tabnote}
\end{table*}

\begin{figure*}
\begin{center} 
 \includegraphics[width=18cm]{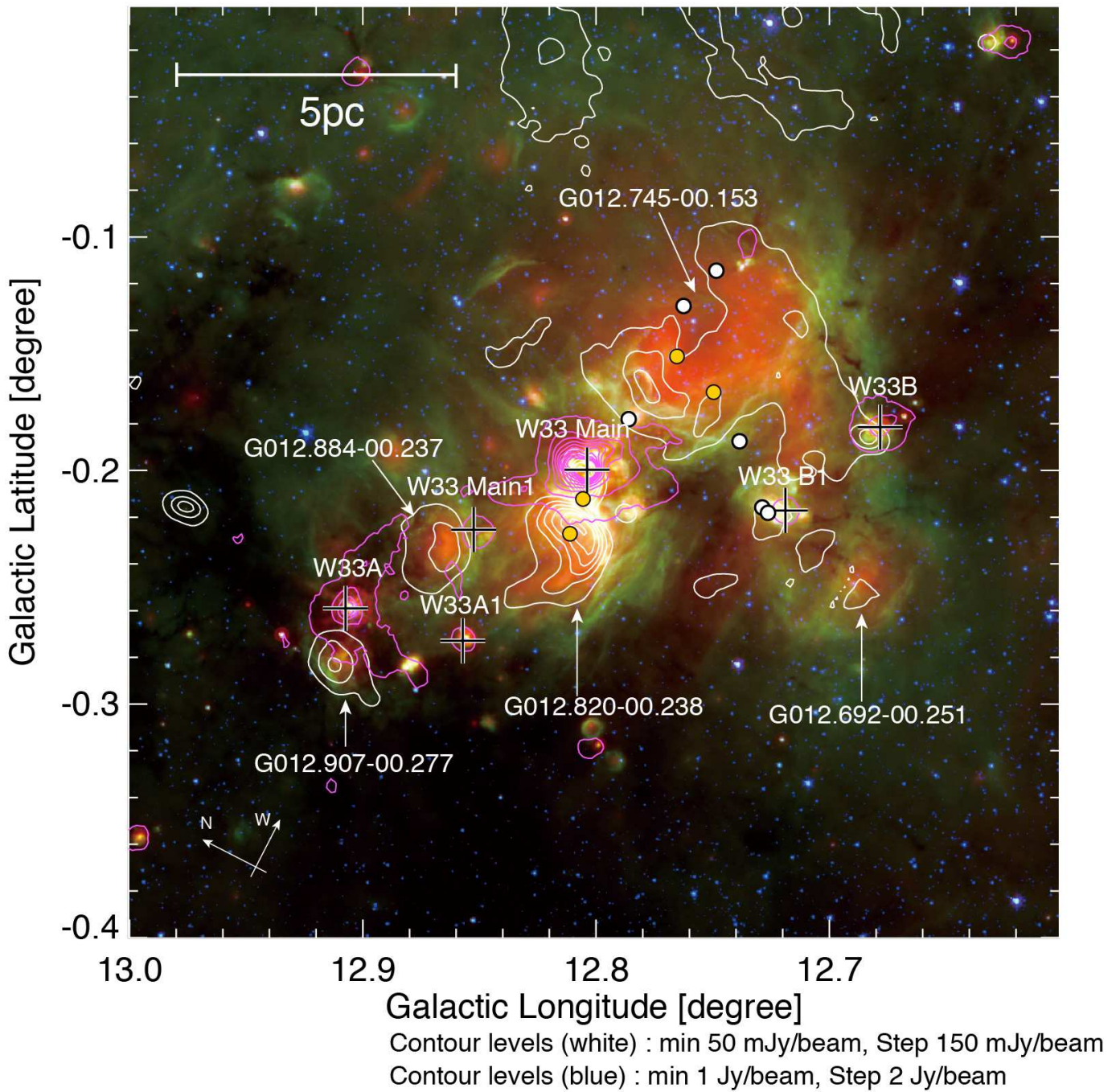}
\end{center}
\caption{Three color image of W33 with blue, green, and red corresponding to {\it Spitzer}/IRAC 3.6 $\mu$m (Benjamin et al. 2003), {\it Spitzer}/IRAC 8 $\mu$m (Benjamin et al. 2003), and {\it Spitzer}/MIPS 24 $\mu$m (Carey et al. 2009) respectively. Large black crosses indicate the { dust clumps} (W33 Main, W33 A, W33 B, W33 Main1, W33 A1, and W33 B1) identified with APEX in 870 $\mu$m by Contreras et al. (2013), and Urquhart et al. (2014). Orange and white circles indicate O and B-type star identified by Messineo et al. (2015). White and { pink} contours show the VLA 90 cm and ATLASGAL 870 $\mu$m continuum image. White arrows show H\,\emissiontype{II} regions identified with the WISE satelite (Anderson et al. 2014, 2015).}\label{.....}
\end{figure*}

\begin{figure*}
\begin{center}  \includegraphics[width=18cm]{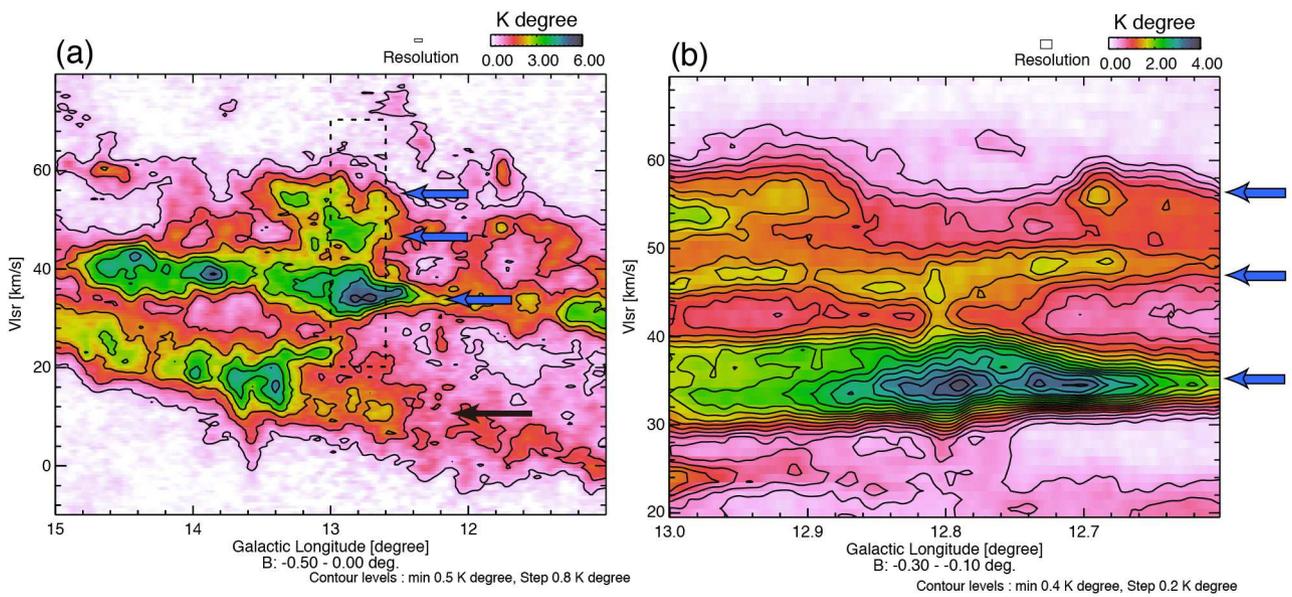}
\end{center}
\caption{(a) Longitude-velocity diagram of the $^{12}$CO$J=$ 1--0 emission with the NANTEN2 telescope. The arrows indicate the four velocity components ; black arrow : 10 km s$^{-1}$, blue arrows : 35 km s$^{-1}$, 45 km s$^{-1}$, and 58 km s$^{-1}$. { The dashed box { indicate} the W33 region as the range of Figure 2(b)}.  
(b) Longitude-velocity diagram of the $^{12}$CO$J=$1--0 emission with the Nobeyama 45-m telescope. The arrows indicate the three velocity components ; 35 km s$^{-1}$, 45 km s$^{-1}$, and 58 km s$^{-1}$.}\label{.....}
\end{figure*}


\clearpage
\begin{figure*}[h]
\begin{center}  \includegraphics[width=17cm]{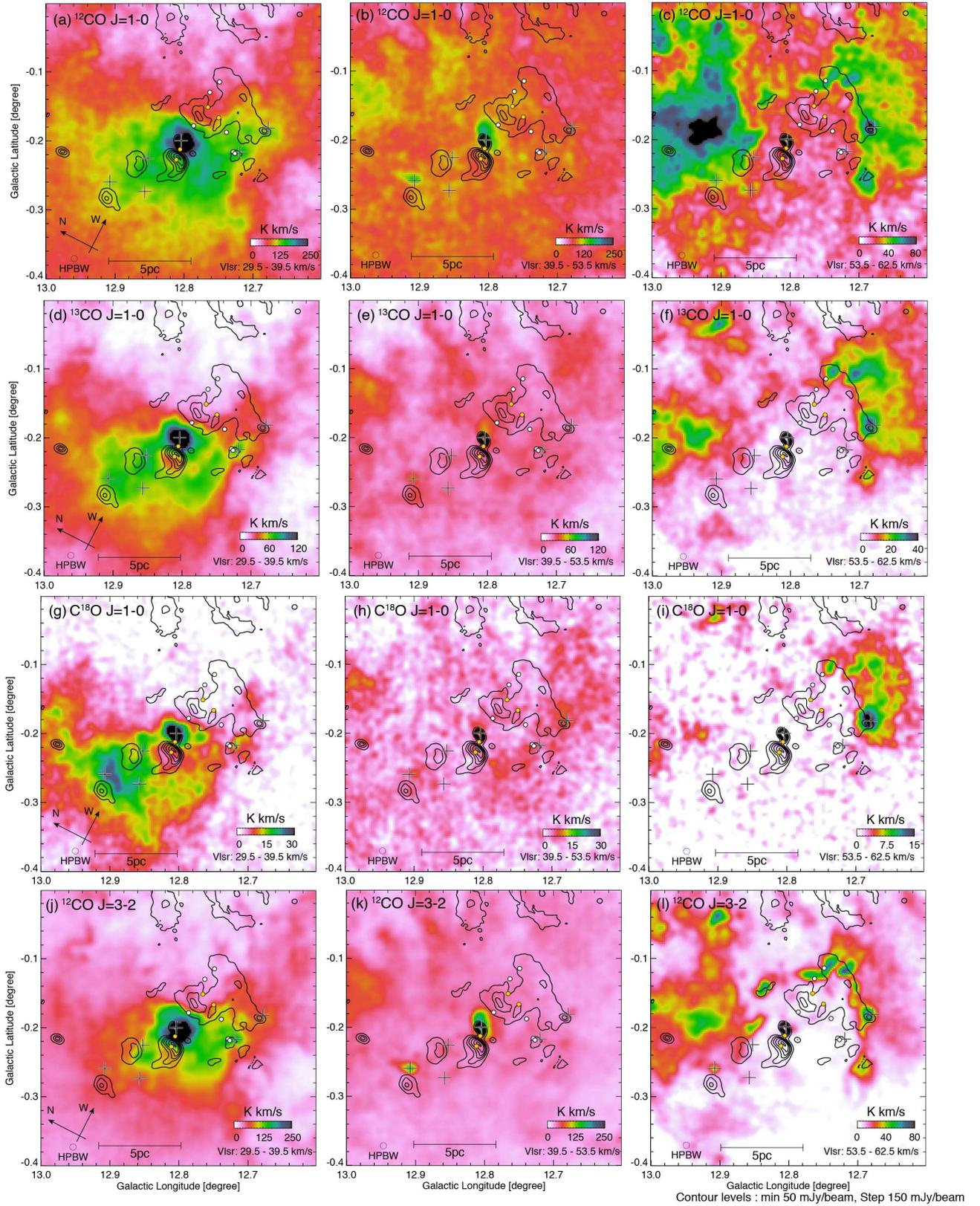}
\end{center}
\caption{Integrated intensity distributions of the 35 km s$^{-1}$ (Left column), 45 km s$^{-1}$ (Center column), and 58 km s$^{-1}$ clouds (Right column).  (a)--(c) $^{12}$CO$J=$1--0, (d)--(f) $^{13}$CO$J=$1--0, and (g)--(i) C$^{18}$O$J=$1--0 are obtained with Nobeyama. (j)--(l) $^{12}$CO$J=$ 3--2 is obtained with JCMT. Contours show the VLA 90 cm radio continuum image. Plots are same as Figure 1.}\label{.....}
\end{figure*}

\begin{figure*}
\begin{center}  \includegraphics[width=14cm]{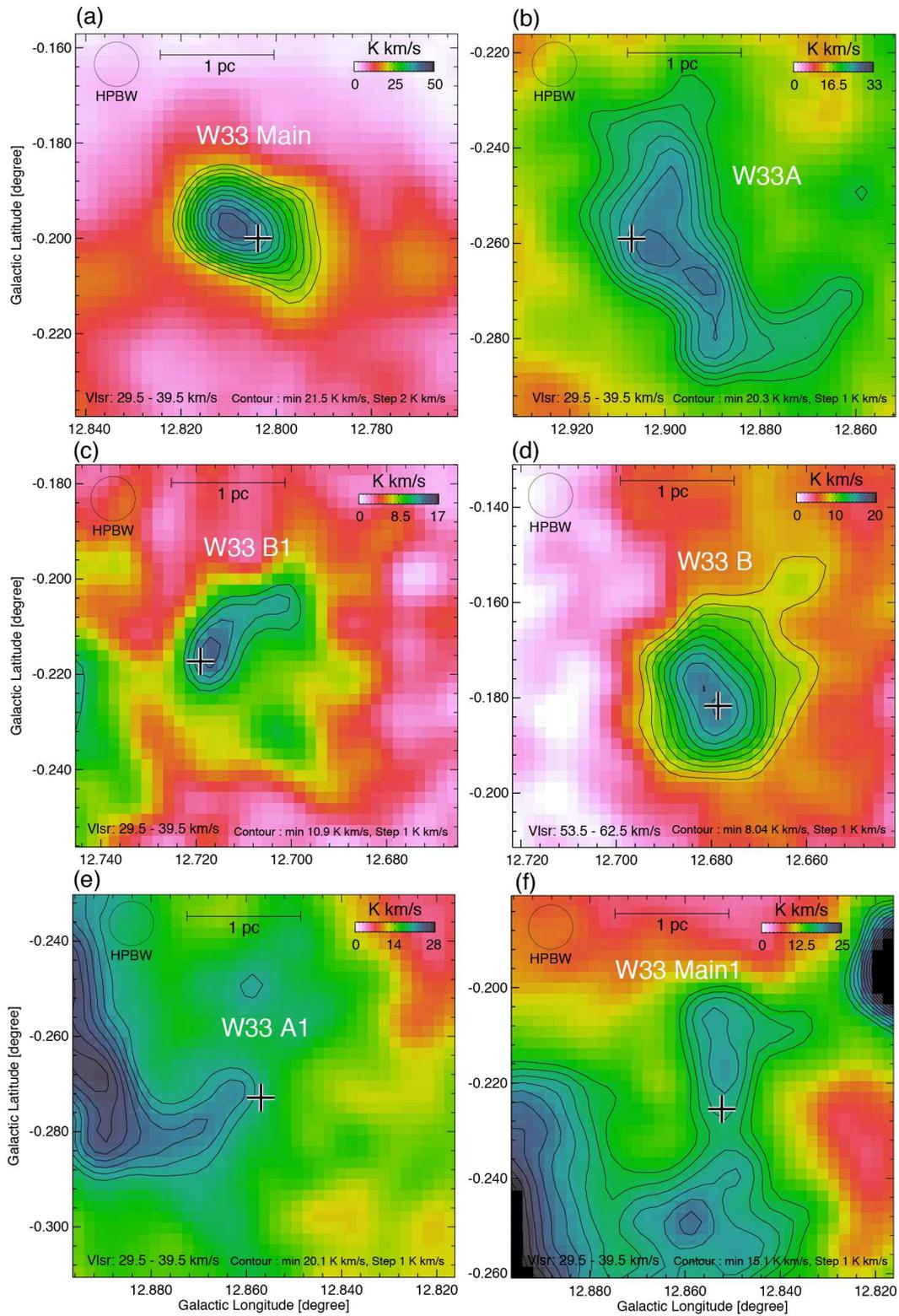}
\end{center}
\caption{C$^{18}$O$J=1-0$ integrated intensity distributions of the molecular clumps by the Nobeyama 45-m telescope. Crosses indicate the dust clumps (W33 Main, W33 A, W33 B, W33 Main1, W33 A1, W33 B1) identified at APEX 870 $\mu$m by Contreras et al. (2013), Urquhart et al. (2014). The lowest contour is defined the half (W33 Main, W33 B) and 75 \% (W33 A, W33 B1, W33 A1, W33 Main1) level of peak integrated intensity. {{ The integrated velocity range shows in the lower left of the figures.} }}\label{.....}
\end{figure*}

\begin{figure*}[h]
\includegraphics[width=16.5cm]{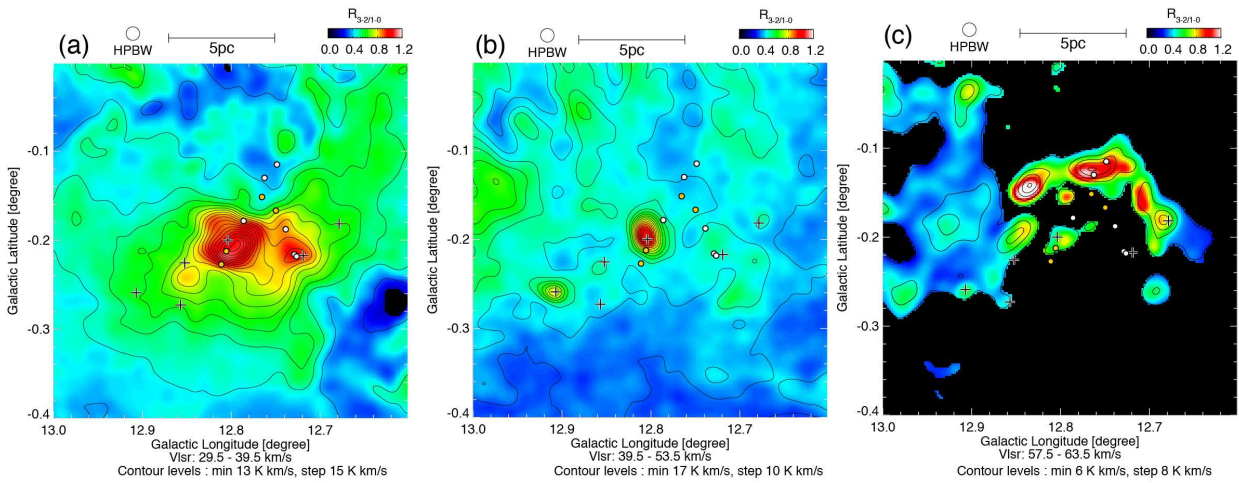}
\caption{$R_{3-2/1-0}$ maps of the (a) 35 km s$^{-1}$, (b) 45 km s$^{-1}$, and (c) 58 km s$^{-1}$ cloud are shown in the color image. Contours show the $^{12}$CO$J=$3--2 emission, which was spatially smoothed \timeform{50"}. Plots are same as Figure 1. { The clipping level is $ 8 \sigma$.} }\label{.....}
\end{figure*}

\begin{figure*}[h]
\begin{center}  \includegraphics[width=17cm]{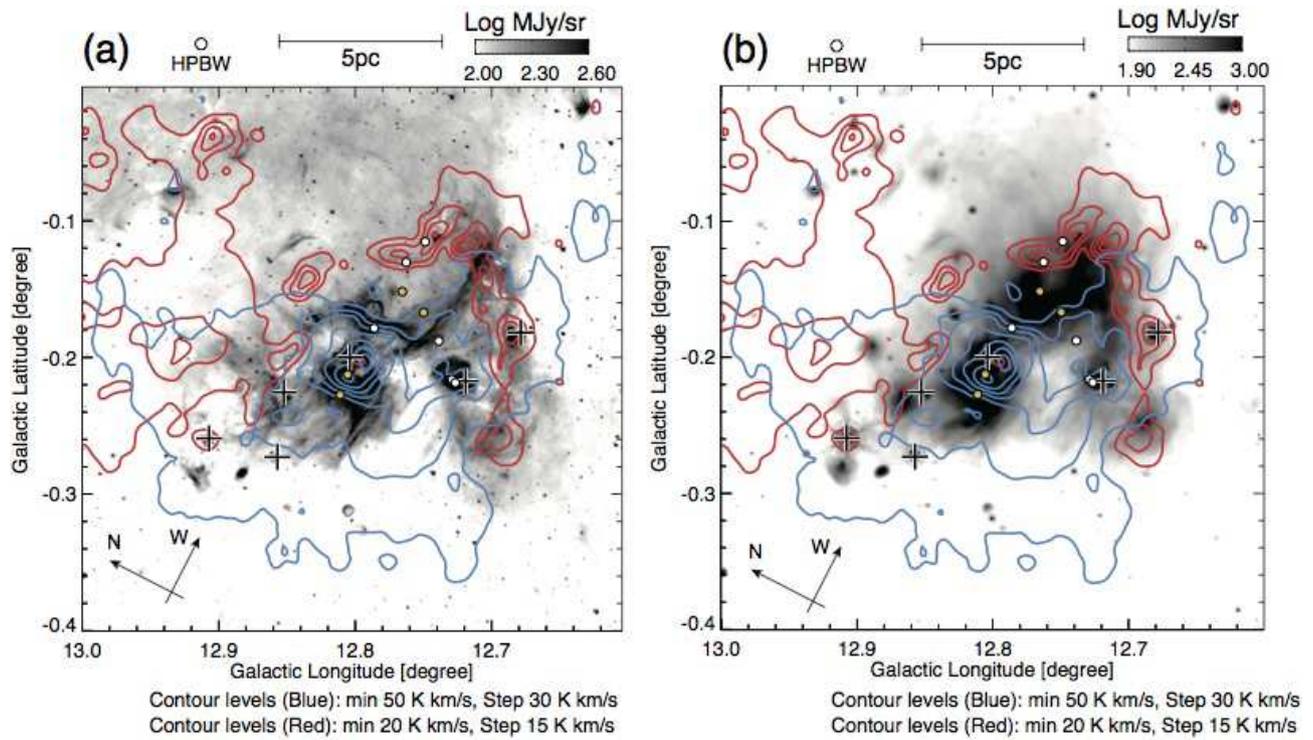}
\end{center}
\caption{Comparison of the {35 km s$^{-1}$ (blue) and 58 km s$^{-1}$ (red)} clouds ($^{12}$CO $J=3-2$ contours) with (a) the {\it Spitzer} 8 $\mu$m and (b) 24 $\mu$m image (grey scale). Plots are same as Figure 1.}\label{.....}
\end{figure*}

\begin{figure*}[h]
\begin{center}  \includegraphics[width=18cm]{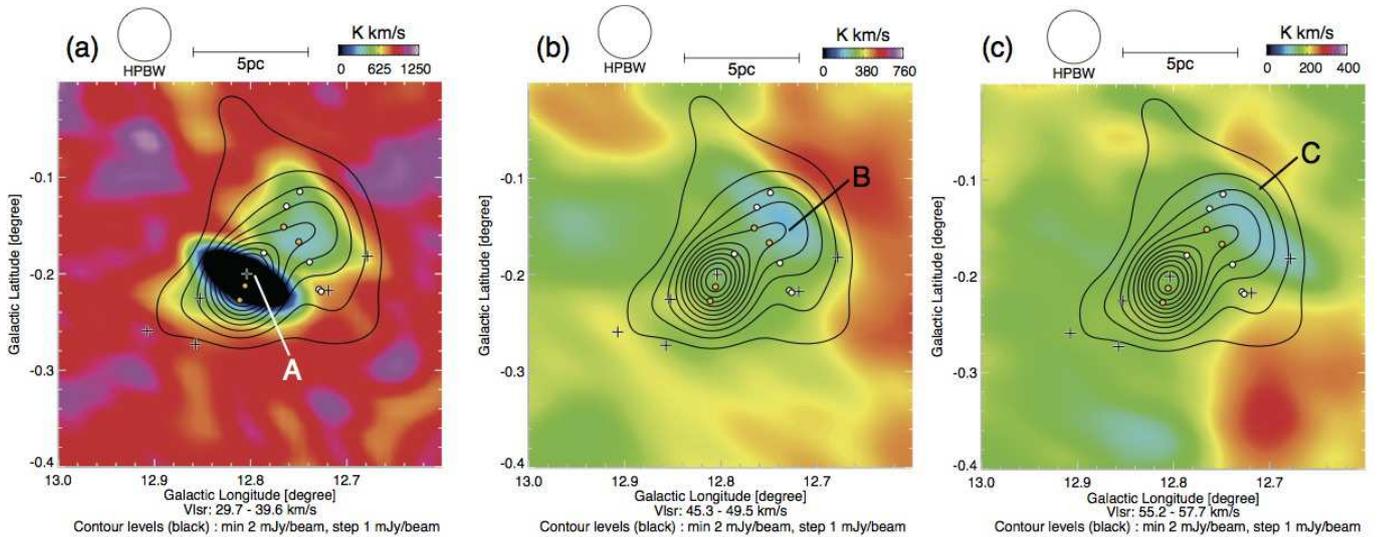}
\end{center}
\caption{Comparison of the H\,\emissiontype{I} 21 cm emission with velocity integration ranges of (a) 30 -- 40 km s$^{-1}$, (b) 45 -- 50 km s$^{-1}$, and (c) 55 -- 58 km s$^{-1}$, respectively and the VLA 20 cm continuum image (Black contour). A, B and C are the positions of spectra in Figure 8 at {$(l,b)=$(\timeform{12.795D},\timeform{-0.202D})}, {(\timeform{12.731D}, \timeform{-0.154D})}, and {(\timeform{12.710D}, \timeform{-0.112D})}, respectively. Plots are same as Figure 1.} \label{.....}
\end{figure*}

\begin{figure*}[h]
\begin{center}  \includegraphics[width=18cm]{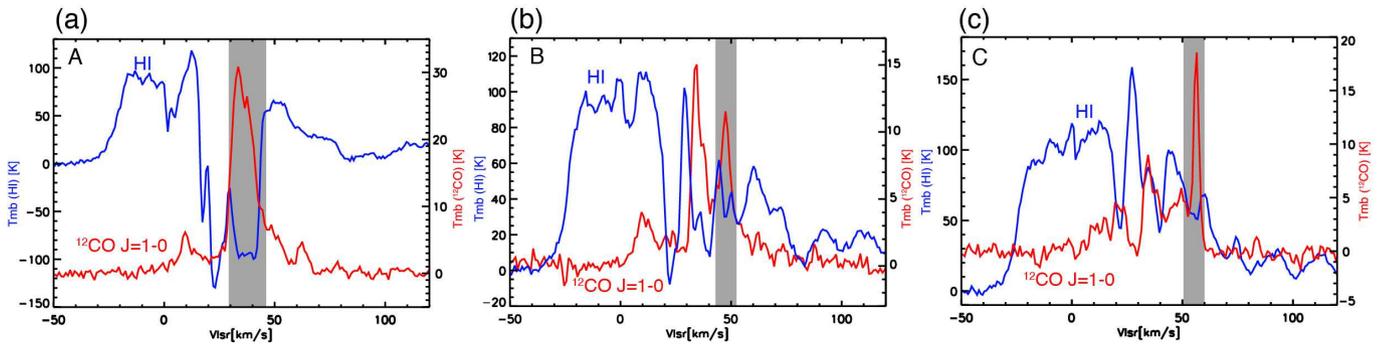}
\end{center}
\caption{The H\,\emissiontype{I} 21 cm and $^{12}$CO $J=$ 1--0 spectrum of  A, B and C in Figure 7 at {$(l,b)=$(\timeform{12.795D},\timeform{-0.202D})}, {(\timeform{12.731D}, \timeform{-0.154D})}, and {(\timeform{12.710D}, \timeform{-0.112D})}, respectively. The gray areas show absorption position of H\,\emissiontype{I} 21 cm emission.} \label{.....}
\end{figure*}

\begin{figure*}
\begin{center}  \includegraphics[width=17cm]{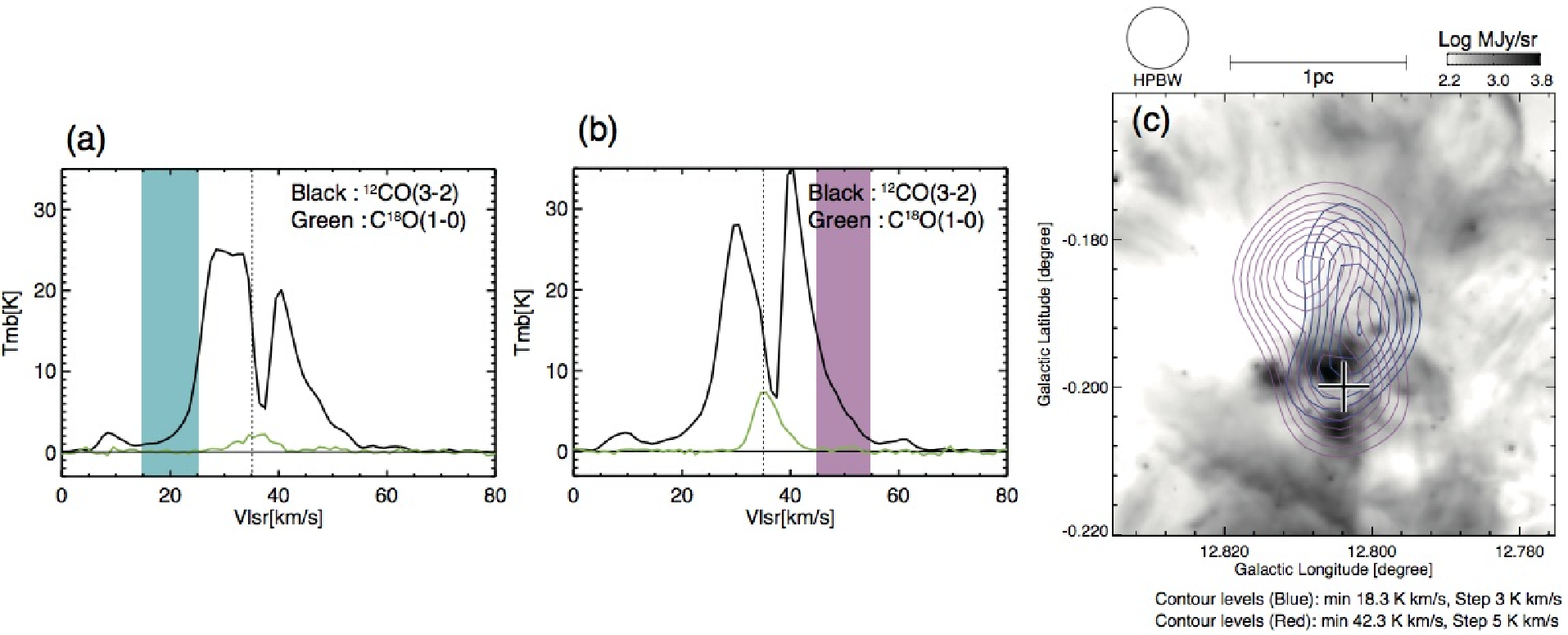}
\end{center}
\caption{(a), (b) The CO spectra of the blue-shifted and the red-shifted lobes obtained at {$(l,b)=$(\timeform{12.800D},\timeform{-0.185D})} and {(\timeform{12.807D}, \timeform{-0.197D})}, respectively. The blue and red area show integrated velocity ranges for the lobes. The dashed lines show systematic velocity. (c) $^{12}$CO$J=$ 3--2 distributions of the molecular outflows associated with W33 Main, where the blue-shifted lobe is shown in blue contours and red-shifted lobe is shown in red contours. The background image is the {\it Spitzer} 8 $\mu$m emission. The cross indicates W33 Main. }\label{.....}
\end{figure*}

\begin{figure*}
\begin{center}  \includegraphics[width=17cm]{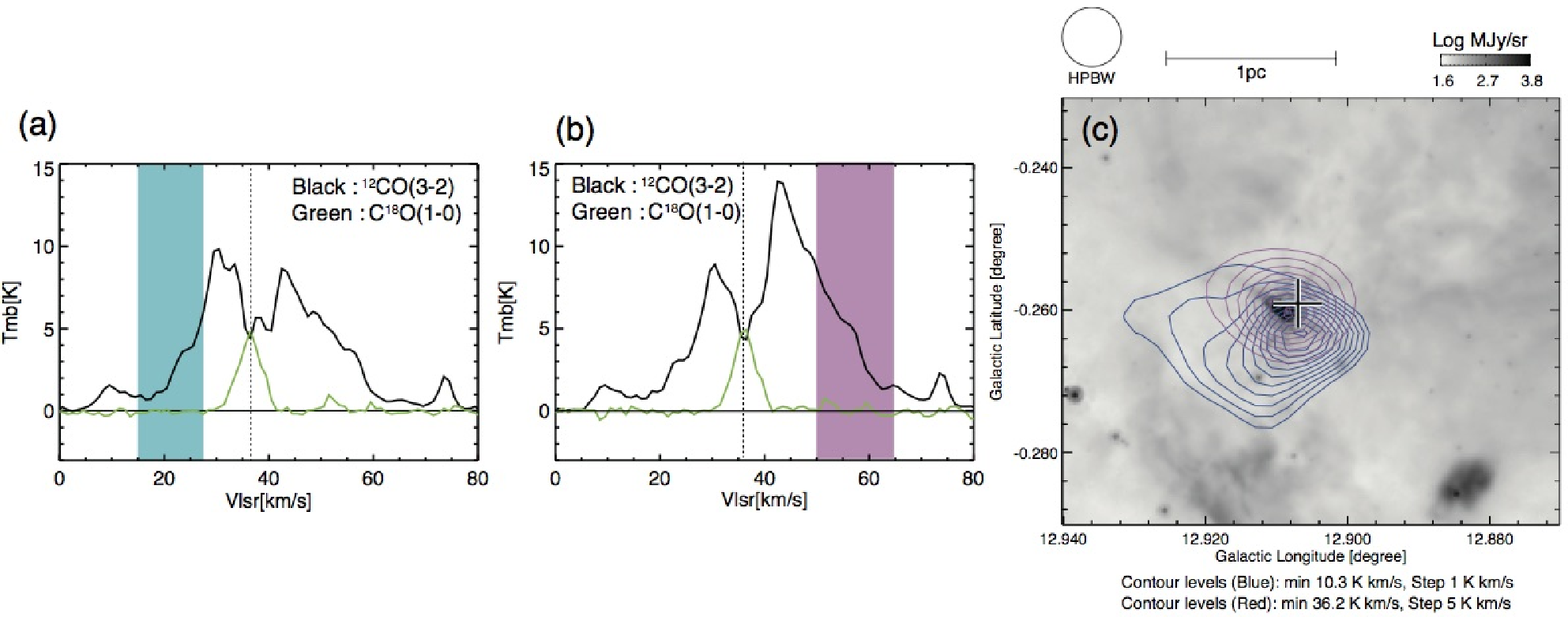}
\end{center}
\caption{(a), (b) The CO spectra of the blue-shifted and the red-shifted lobes obtained at {$(l,b)=$(\timeform{12.906D}, \timeform{-0.266D})} and {$(l,b)=$(\timeform{12.908D}, \timeform{-0.261D})}, respectively. The blue and red area show integrated velocity ranges for the lobes. The dashed lines show systematic velocity. (c) $^{12}$CO$J=$ 3--2 distributions of the molecular outflows associated with W33 A, where the blue-shifted and red-shifted lobes are shown in blue and red respectively. The background image is the {\it Spitzer} 8 $\mu$m emission. The cross indicates W33 A.}\label{.....}
\end{figure*}

\clearpage


\begin{figure*}[h]
\begin{center} 
 \includegraphics[width=10.5cm]{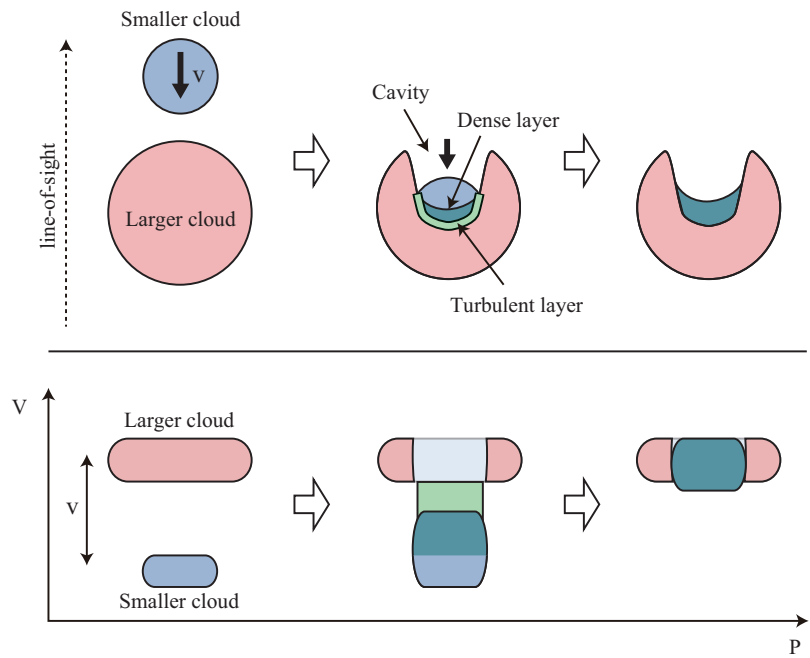}
\end{center}
\caption{Schematic of a collision between two dissimilar clouds and cartoons position-velocity diagrams, where the gas density in the smaller cloud is much smaller than that in the larger cloud. Different color components in the collision schematics correspond to the different color on the position-velocity diagrams. }\label{.....}
\end{figure*}

\begin{figure*}[h]
\begin{center} 
 \includegraphics[width=10.5cm]{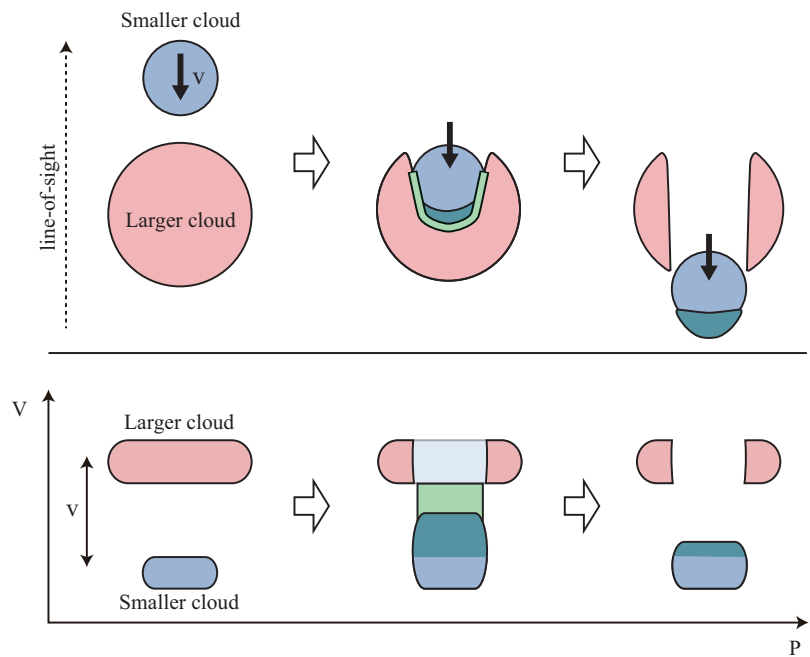}
\end{center}
\caption{Same as Figure 11 but for the case that the gas density in the smaller cloud is much higher than that in the larger cloud }\label{.....}
\end{figure*}

\begin{figure*}[h]
\begin{center} 
 \includegraphics[width=18cm]{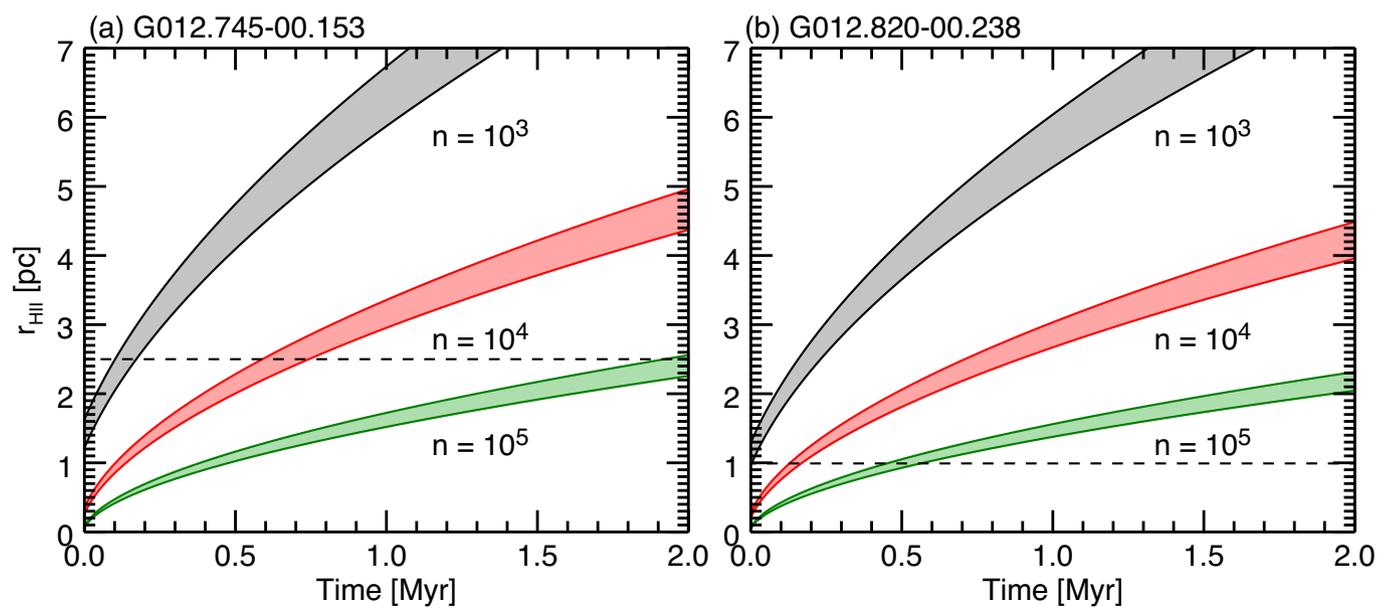}
\end{center}
\caption{Evolutionary tracks of expanding the H\,\emissiontype{II} region radius $r_{\rm H\,\emissiontype{II}}$ are plotted for different initial densities, based on the analytical model of the D-type expansion by Spitzer (1978). (a) and (b) show G012.745-00.153 and G012.820-00.238, respectively. Dashed lines show the radius of H\,\emissiontype{II} regions.}\label{.....}
\end{figure*}

\clearpage
\appendix
\section*{CO velocity channel maps of  the Nobeyama and JCMT data sets}
We show the velocity channel maps of the $^{12}$CO, $^{13}$CO, C $^{18}$O $J=$1--0, $^{12}$CO $J=$3--2 emission, and $^{12}$CO $J=$1--0/$J=$3--2 ratio of W33 in Figure 14--18, respectively. The velocity range is between 27.5 and 63.5 km s$^{-1}$. The contours indicate the 90 cm radio continuum emission with VLA.

\begin{figure*}[h]
 \includegraphics[width=18cm]{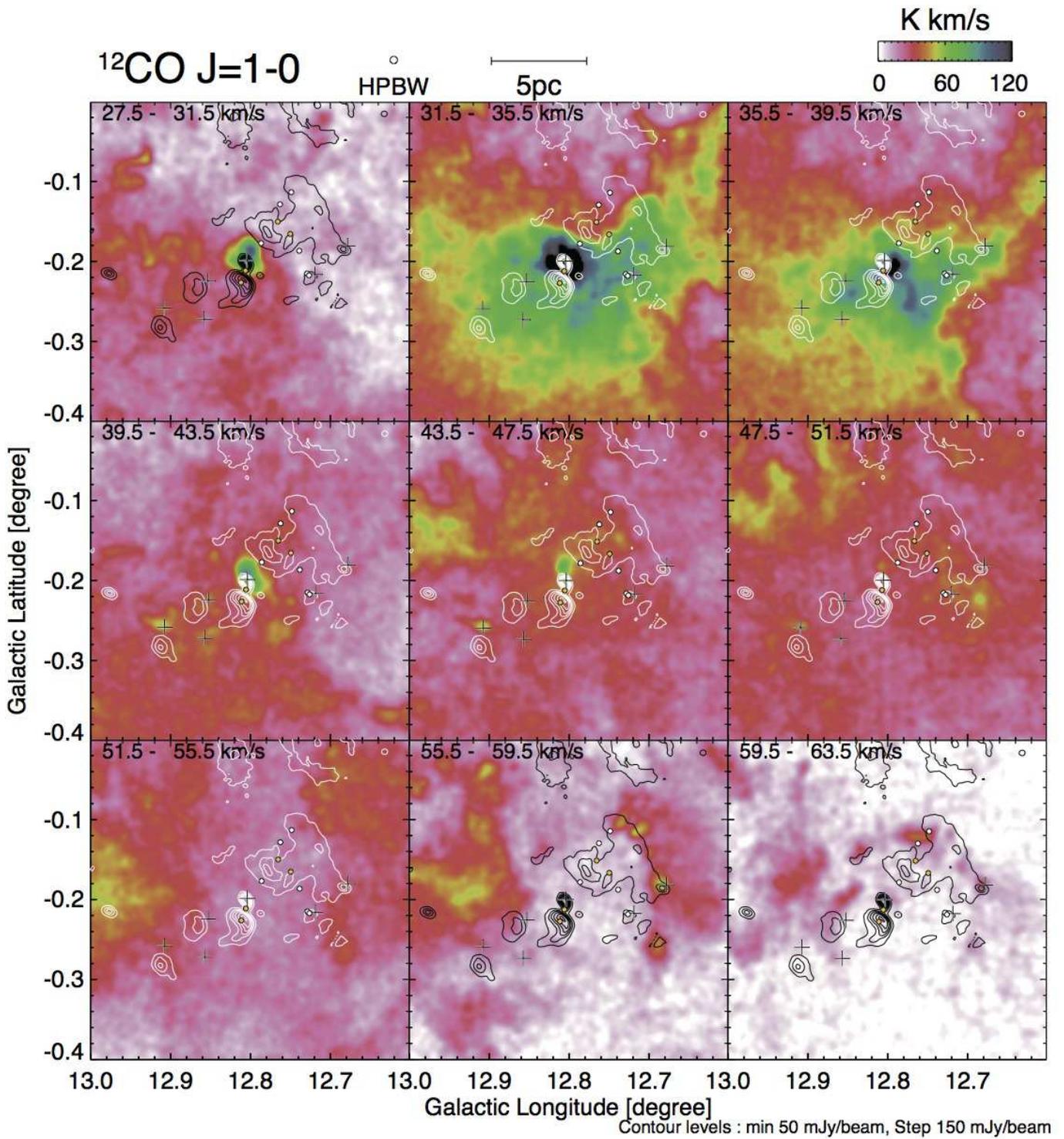}
\caption{Velocity channel map of the $^{12}$CO $J=$ 1--0 emission with a velocity step of  4.0 km s$^{-1}$ with Nobeyama. Contours show the VLA 90 cm radio continuum image. Plots are same as Figure 1.}\label{.....}
\end{figure*}

\begin{figure*}[h]
 \includegraphics[width=18cm]{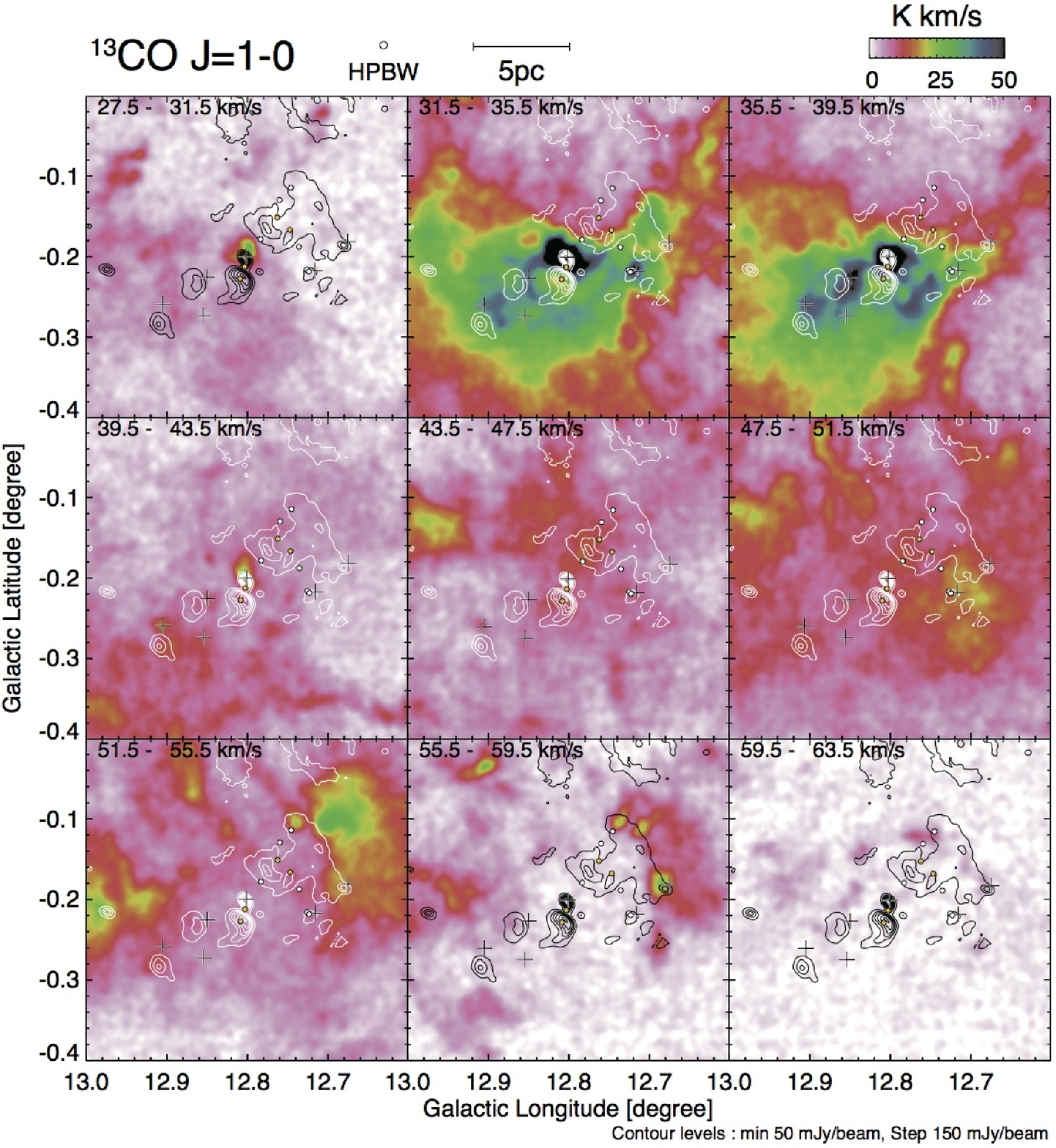}
\caption{Velocity channel map of the $^{13}$CO $J=$ 1--0 emission with a velocity step of  4.0 km s$^{-1}$ with Nobeyama. Contours show the VLA 90 cm radio continuum image. Plots are same as Figure 1. }\label{.....}
\end{figure*}

\begin{figure*}[h]
 \includegraphics[width=18cm]{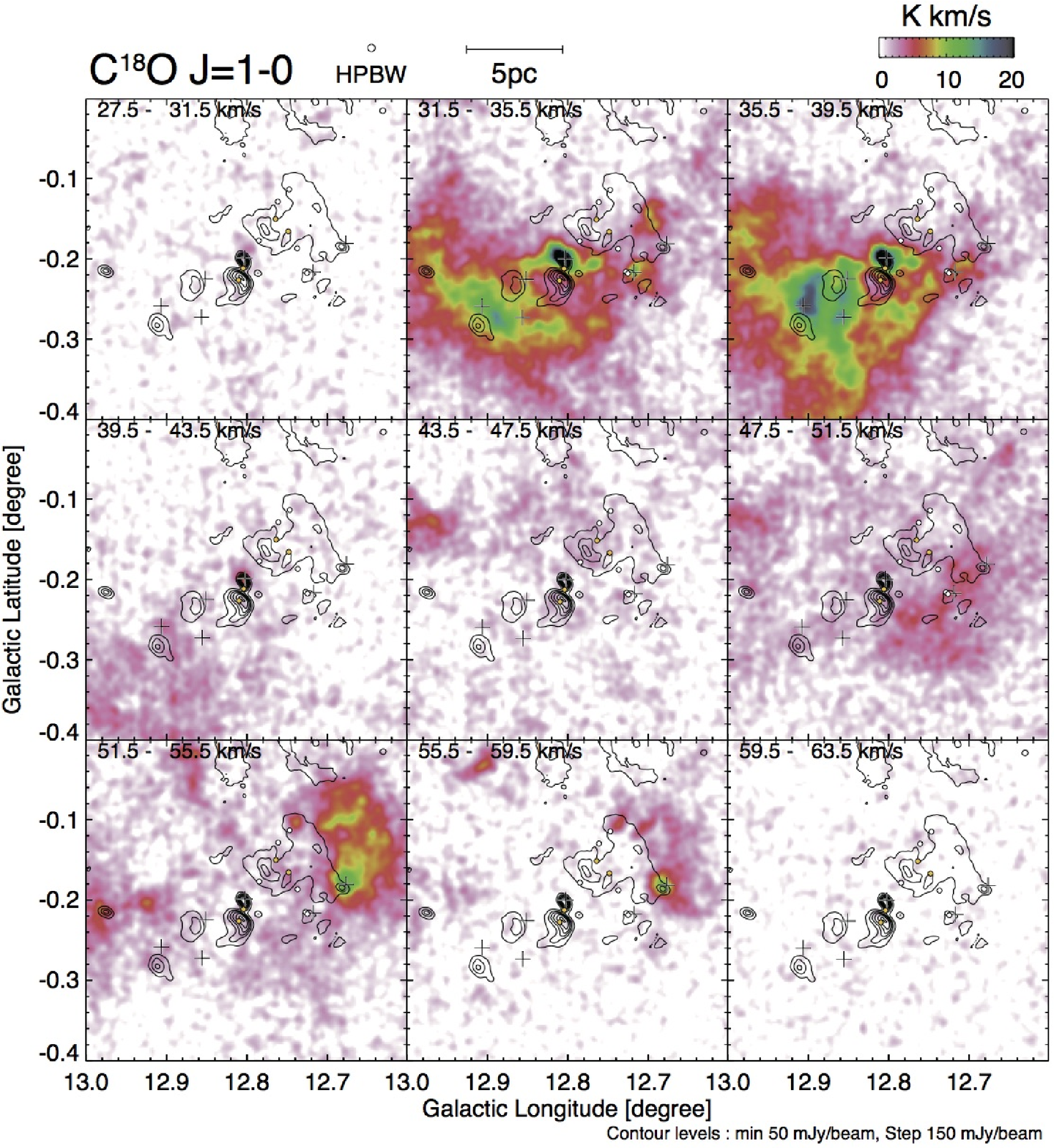}
\caption{Velocity channel map of the C$^{18}$O $J=$ 1--0 emission with a velocity step of  4.0 km s$^{-1}$ with Nobeyama. Contours show the VLA 90 cm radio continuum image. Plots are same as Figure 1. }\label{.....}
\end{figure*}

\begin{figure*}[h]
 \includegraphics[width=18cm]{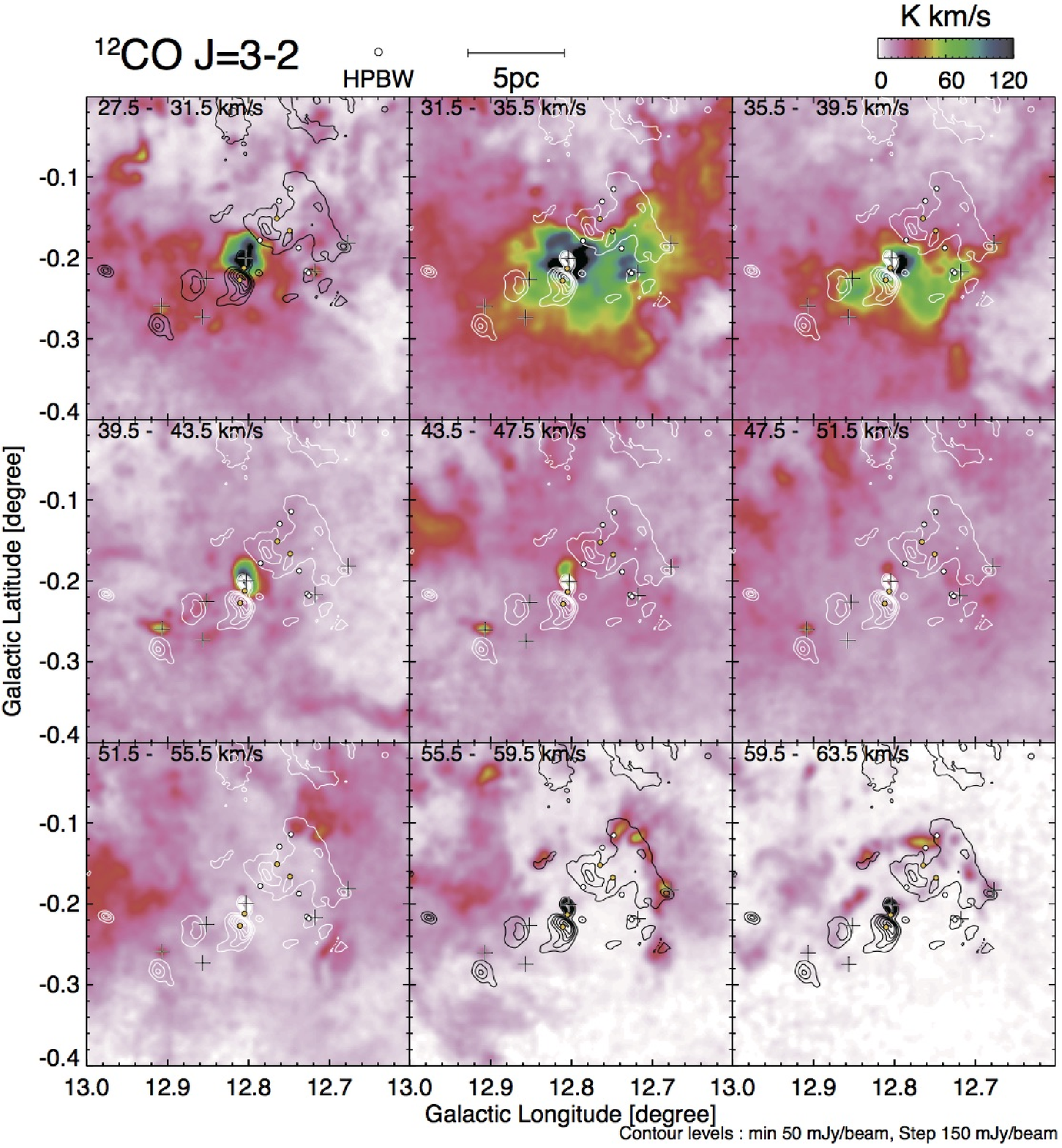}
\caption{Velocity channel map of the $^{12}$CO $J=$ 3--2 emission with a velocity step of  4.0 km s$^{-1}$ with JCMT. Contours show the VLA 90 cm radio continuum image. Plots are same as Figure 1.}\label{.....}
\end{figure*}

\begin{figure*}[h]
 \includegraphics[width=18cm]{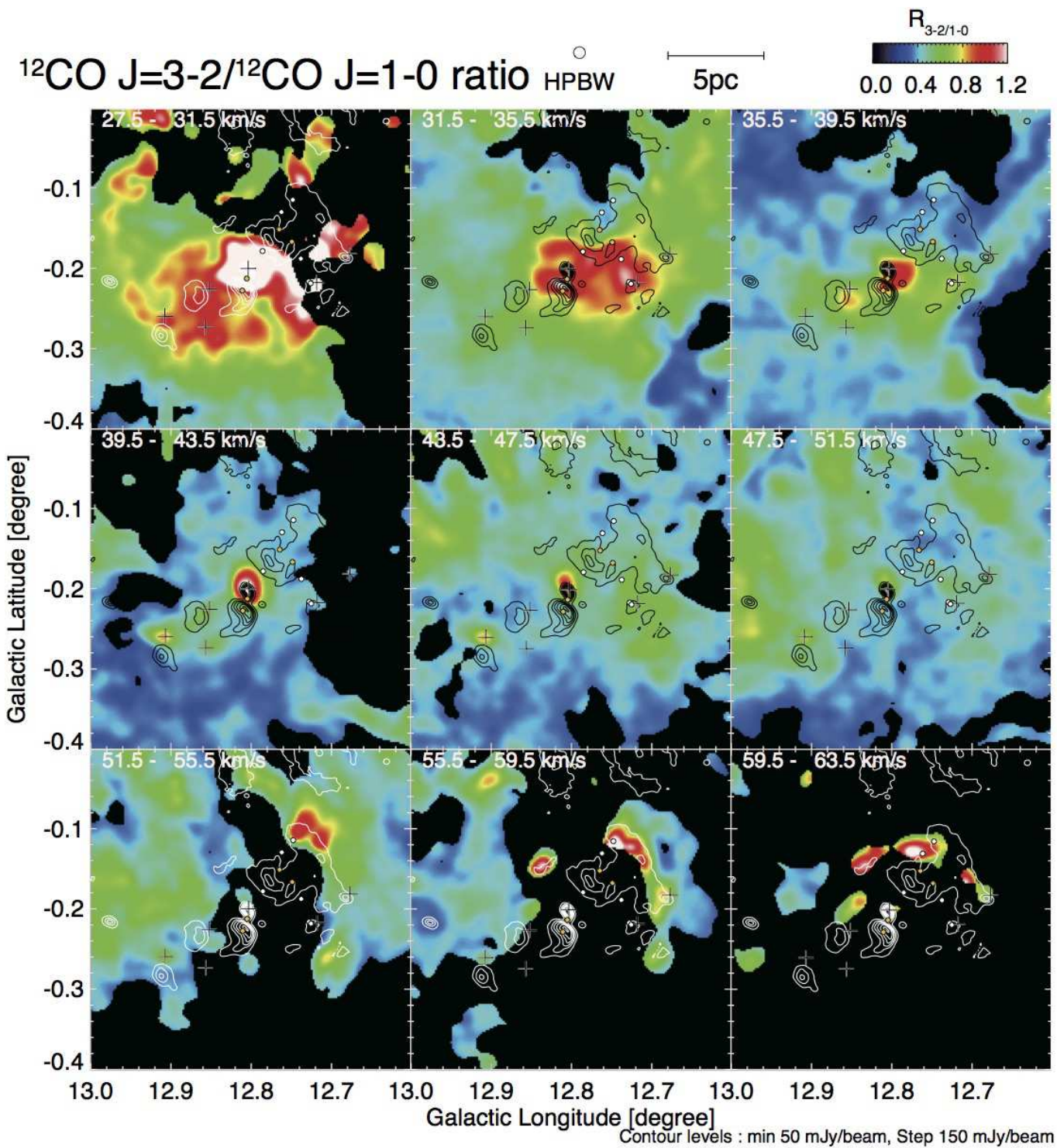}
\caption{Velocity channel map of the $^{12}$CO $J=$ 3--2 /$^{12}$CO $J=$ 1--0 emission with a velocity step of  4.0 km s$^{-1}$, which was spatially smoothed \timeform{50"}. Contours show the VLA 90 cm radio continuum image. Plots are same as Figure 1. { The clipping level is $ 8 \sigma$ (4.2 K km s$^{-1}$).} }\label{.....}
\end{figure*}

\end{document}